\documentclass[a4paper,fleqn,usenatbib]{mnras}

\usepackage{times,txfonts}

\usepackage[T1]{fontenc}
\usepackage{ae,aecompl}
\def\mdot{$\dot{M}$}
\def\teff{$T_{\rm eff}$}

\usepackage{graphicx}	

\title[Evolution of RSGs to supernova]{The evolution of Red Supergiants to supernova in the LMC cluster NGC 2100}

\author[E. R. Beasor \& B. Davies]{
Emma R. Beasor,$^{1}$\thanks{E-mail:e.beasor@2010.ljmu.ac.uk}
Ben Davies$^{1}$
\\
$^{1}$Astrophysics Research Institute, Liverpool John Moores University, Liverpool,  L3 5RF, UK}

\date{Accepted XXX. Received YYY; in original form ZZZ}

\pubyear{2016}

\begin{document}
\label{firstpage}
\pagerange{\pageref{firstpage}--\pageref{lastpage}}
\maketitle

\begin{abstract}
The mass loss rates of red supergiants (RSGs) govern their evolution towards supernova and dictate the appearance of the resulting explosion. To study how mass-loss rates change with evolution we measure the mass-loss rates (\mdot) and extinctions of 19 red supergiants in the young massive cluster NGC2100 in the Large Magellanic Cloud. By targeting stars in a coeval cluster we can study the mass-loss rate evolution whilst keeping the variables of mass and metallicity fixed. Mass-loss rates were determined by fitting DUSTY models to mid-IR photometry from WISE and Spitzer/IRAC. We find that the \mdot\ in red supergiants increases as the star evolves, and is well described by \mdot\ prescription of de Jager, used widely in stellar evolution calculations. We find the extinction caused by the warm dust is negligible, meaning the warm circumstellar material of the inner wind cannot explain the higher levels of extinction found in the RSGs compared to other cluster stars. We discuss the implications of this work in terms of supernova progenitors and stellar evolution theory. We argue there is little justification for substantially increasing the \mdot\ during the RSG phase, as has been suggested recently in order to explain the absence of high mass Type IIP supernova progenitors. We also argue that an increase in reddening towards the end of the RSG phase, as observed for the two most evolved cluster stars, may provide a solution to the red supergiant problem. 
\end{abstract}




\section{Introduction}
Knowledge of the mass loss rates (\mdot) of red supergiants (RSGs) is fundamentally important for understanding stellar evolution. Changing \mdot\ has effects on the subsequent evolution of the star, as well as the supernova (SN) type and eventual remnant \citep[e.g.][]{maeder1981most,chiosi1986evolution}.

When a RSG reaches the end of it's lifetime, it explodes as a Type IIP core-collapse supernova (CCSN), of which there have been 7 confirmed cases of RSGs as progenitors, the most recent being the 12.5 $\pm$ 1.2 M$_{\odot}$ progenitor to SN 2012aw \citep{fraser2016disappearance}.Theory predicts that the RSG progenitor stars of Type IIP supernovae can be anywhere in the range of 8.5 to 25 M$_{\odot}$ \citep[e.g.][]{meynet2003stellar},  but so far it seems the stars which explode are of a relatively low mass (15M$_{\odot}$), with no progenitors appearing in the higher end of the predicted mass range \citep[between 17 and 25 M$_{\odot}$][]{smartt2009death,smartt2015observational}. 

Are all RSGs exploding as Type IIP SNe? Or does the extreme mass loss affect the final evolution of these massive stars? Stellar evolution models currently rely on observational or theoretical mass loss rate prescriptions \citep[e.g.][]{de1988mass,reimers1975circumstellar,van2005empirical,nieuwenhuijzen1990parametrization,feast1992ch}. A potential weakness of these prescriptions is that they have relied on observations of field stars, not coeval stars, leaving parameters of initial mass ($M_{\rm initial}$) and metallicity ($Z$) unconstrained which could potentially explain the large dispersions in the observed trends. \cite{georgy2015mass} discuss the implications of extreme mass loss that could occur at the end of RSGs lives \citep[see also][]{georgy2012yellow}. In this paper it was found that high mass loss rates of RSGs can lead to a blueward movement in the Hertzsprung-Russel diagram (HRD), occuring for RSGs more massive than 25M$_{\odot}$ (non-rotating models) and 20M$_{\odot}$ (rotating models). \citet{georgy2015mass} find that this blueward motion allows to fit the observed maximum mass of observed type IIP SNe progenitors. They also express the need for better determination of RSG mass-loss rates to improve the stellar modeling at this evolved stage of the star's life. 

In addition to the effect on RSG evolution, it has been claimed that circumstellar dust in the wind could, in part, provide a solution to the missing high mass RSG progenitors. It is known that RSG form dust in their winds \citep[e.g][]{de2008red} and infrared interferometry has shown that this dust can lie very close to the star itself \citep{danchi1994characteristics}. \cite{walmswell2012circumstellar} have shown that by failing to take into account the additional extinction resulting from RSG winds, the luminosity of the most massive red supergiants at the end of their lives is underestimated. Mass estimates are based on mass-luminosity relations meaning that extra intrinsic extinction close to RSG progenitors would give reduced luminosities. While \cite{smartt2009death} did provide extinction estimates of nearby supergiants and of the SN itself throughout their paper, it is suggested that these could be underestimates \citep{walmswell2012circumstellar}. It is expected that this dust would then be destroyed in the SN explosion and hence not be seen in the SN spectra. 

In this paper we measure the amount of circumstellar material and estimate mass-loss rates, to investigate whether this is correlated with how close the star is to supernova. We model the mid-IR excess of 19 RSGs in stellar cluster NGC2100, each which we assume has the same initial mass and composition, but where the stars are all at slightly different stages of evolution. This allows us to investigate the \mdot\ behaviour with evolution of the RSG.

 We begin in Section 2 by describing our dust shell models and choice of input parameters. In Section 3 we discuss applying this to the stars in cluster NGC 2100 and the results we derive from our models. in Section 4 we discuss our results in terms of RSG evolution and as progenitors. 

\section{Dust shell models}
The models used in this project were created using DUSTY \citep{ivezic1999dusty}. Stars surrounded by circumstellar dust have their radiation absorbed/re-emitted by the dust particles, changing the output spectrum of the star. DUSTY solves the radiative transfer equation for a star obscured by a spherical dust shell of a certain optical depth ($\tau_V$, optical depth at 0.55 $\micron$), inner dust temperature ($T_{\rm in}$) at the inner most radius ($R_{\rm in}$). Below we describe our choices for the model input parameters and our fitting methodology.

\subsection{Model parameters}
\subsubsection{Dust composition}
It is necessary to define a dust grain composition when creating models with DUSTY as this determines the extinction efficiency Q$_\lambda$, and hence how the dust shell will reprocess the input spectral energy distribution (SED). Observations of RSGs confirm the dust shells are O-rich, indicated by the presence of mid-IR spectral features at 12 and 18$\micron$ known to be caused by the presence of silicates. We opted for O-rich silicate dust as described by \cite{draine1984optical}. Ossenkopff 'warm' and 'cold' silicates \citep{ossenkopf1992constraints} were also considered resulting in only small changes to the output flux. The difference in fluxes from each O-rich dust type were found to be smaller than the errors on our photometry. We therefore concluded that our final results were insensitive to which O-rich dust type we chose.
\subsubsection{Grain size, a}
DUSTY also requires a grain size distribution to be specified. \cite{kochanek2012absorption} used DUSTY to model the spectrum for the RSG progenitor of SN 2012aw, opting for the MRN power law with sharp boundaries, \citep[$dn/da \propto a^{-3.5}$ for 0.005  $\micron$ $<$ a $<$ 0.25  $\micron$ ][]{mathis1977size}. This power law is more commonly associated with dust grains in the interstellar medium. \cite{van2005empirical} also used DUSTY to model dust enshrouded RSGs, choosing a constant grain size of 0.1$\micron$. However, it is also stated in this paper that the extinction of some of the most dust enshrouded M-type stars was better modelled when a smaller grain size, 0.06$\micron$ was used or a modified MRN distribution, between 0.01 and 0.1$\micron$. \cite{groenewegen2009luminosities} also investigated the effect of grain size on the output spectrum, finding a grain size of 1$\micron$ fit reasonably well to the observations of O-rich RSG stars in the SMC and LMC. Recent observations of VY Canis Majoris \citep{scicluna2015large}, a nearby dust-enshrouded RSG, estimated the dust surrounding the star to be of a constant grain size of 0.5$\micron$. This is in line with previous observations such as those by \cite{smith2001asymmetric}, who found the grain size to be between 0.3 and 1$\micron$. Taking all this into account we created models for the MRN power law as well as constant grain sizes of 0.1, 0.2, 0.3, 0.4 and 0.5$\micron$, choosing 0.3$\micron$ as our fiducial grain size. However, as we are studying stars' emission at wavelengths much greater than grain size ($\lambda $> 3$\micron$) the scattering and absorption efficiencies of the dust is largely independent of the grain size. This is discussed further in Section 3.4.    
\subsubsection{Density distribution}           
Here, we assumed a steady state density distribution falling off as $r^{-2}$ in the entire shell with a constant terminal velocity. We do not know the outflow velocity for RSGs in our sample so we rely on previous measurements to estimate this value. \cite{van2001circumstellar} and \cite{richards1998maser} both used maser emission to map the dust shells of other RSGs,finding v$_{\infty}$ values consistent with $\sim$20-30 km/s for the stars in their samples. We opted for a uniform rate of 25 $\pm$ 5 km/s for the outflow wind. Radiatively driven wind theory suggests that $v_{\infty}$ scales with luminosity, $v_\infty$ $\propto$ L$^{1/4}$, though this is negligible for the luminosity range we measure compared to our errors on luminosity. We specify that the shell extends to 10000 times its inner radius, such that the dust density is low enough at the outer limit so that it has no effect on the spectrum. We also require an gas to dust ratio to be input, $r_{gd}$. It has been shown that this quantity scales with metallicity \citep{marshall2004asymptotic}, so while the gas to dust ratio for RSGs in the Milky Way is around 1:200, for RSGs in the more metal poor LMC the value is higher, around 1:500. We also assumed a grain bulk density, $\rho_d$ of 3g cm$^{-3}$. The values adopted for $r_{gd}$ and $r_s$ will have an effect on the absolute values of \mdot. It is likely that changes in these properties would have little to no effect on the relative \mdot\ values and the correlation with luminosity, but the absolute value of the relation may change. 

The calculation of \mdot\ requires the calculation of $\tau_\lambda$ between $R_{\rm in}$ and $R_{\rm out}$ 

\begin{equation} \tau_\lambda = \int\limits_{R_{\rm in}}^{R_{\rm out}} \pi a^2 Q_\lambda n(r) dr
\end{equation}
for a certain number density profile, $n(r) = n_0 (R_{\rm in}/r) ^2$, where $n_0$ is the number density at the inner radius, $R_{\rm in}$, and extinction efficiency, $Q_\lambda$. We can rearrange to find the mass-density $\rho_0$ at $R_{\rm in}$, 

\begin{equation}
\rho_0 = \frac{4}{3}\frac{\tau_\lambda \rho_d a}{Q_\lambda R_{in}}
\end{equation}

By substituting this into the mass continuity equation ($\dot{M}=4 \pi r^2 \rho(r) v_\infty$)  a mass loss rate can be calculated, 
 \begin{equation} \dot{M} = \frac{16\pi}{3} \frac{R_{in} \tau_{\lambda}  \rho_d a v_\infty}{Q_{\lambda}}r_{gd}
 \end{equation}

 Our choice of density distribution differs from that used in other similar work, for example \cite{shenoy2016searching}, who performed a similar study on the red supergiants $\mu$ Cep and VY CMa. By adopting a constant $T_{\rm in}$ value of 1000K and allowing the density exponent to vary, \cite{shenoy2016searching} found that the best fits were obtained by adopting exponents < 2, and hence concluded that \mdot\ decreases over the lifetime of the stars. In Section 4 we show that this can be reconciled by fixing the density exponent, q=2, and allowing $T_{\rm in}$ to vary. While 1200K is the commonly adopted temperature for silicate dust sublimation, there are many observations in the literature that suggest dust may begin to form at lower $T_{\rm in}$, and hence larger radii. There is interferometric data supporting the case for RSGs having large dust free cavities, for example \cite{ohnaka2008spatially}, who used $N$-band spectro-interferometric observations to spatially resolve the dust envelope around LMC RSG WOH G64. \cite{sargent2010mass} used radiative transfer models of dust shells around two LMC AGB stars, finding bets fit models with lower $T_{\rm in}$ values of 430K and 900K. These values are comparable to previous determinations of $T_{\rm in}$ for O-rich stars from mid-IR infrared fitting similar to the work presented here \citep[e.g.][]{bedijn1987dust,schutte1989theoretical,van2005empirical} suggesting $T_{\rm in}$ often varies from the hot dust sublimation temperature of 1000-1200K.
  
\subsubsection{Sensitivity to $T_{eff}$}
DUSTY requires an input SED to illuminate the dust shell, so that the light can be reprocessed and re-emitted. The SEDs we use are synthesised from MARCS model atmospheres \citep{gustafsson2008grid} using TURBOSPECTRUM \citep{plez2012turbospectrum}. We opted for typical RSG parameters (log(g)=0.0, microturbulent velocity 4km/s) and an LMC-like metallicity of [Z]=-0.3, though the precise value of these parameters are relatively inconsequential to the morphology of the SED. The most important parameter is the stellar effective temperature, \teff. \cite{patrick2016chemistry} used KMOS spectra of 14 RSGs in NGC2100 (of which 13 are analysed in this paper), finding the average \teff\ to be 3890 $\pm$ 85 K. This is consistent with the temperature range observed for RSGs in the LMC by \cite{davies2013temperatures}, who found the average $T_{\rm eff}$ of a sample of RSGs in the LMC to be 4170 $\pm$ 170K, by using VLT+XSHOOTER data  and fitting this to line-free continuum regions of SEDs. In this study we have opted for a fiducial SED of \teff = 3900K in line with these findings. We also checked how sensitive our results were to this choice of SED temperature by re-running the analysis with stellar SEDs $\pm$ 300K of our fiducial SED, fully encompassing the range observed by \cite{patrick2016chemistry}. We found that the different SEDs reproduced the mid-IR excess, and therefore the inferred \mdot, almost identically with very small errors $<$ 10\%. Different $T_{\rm eff}$ values do however affect the bolometric correction and therefore the $L_{\rm bol}$, leading to errors of $\sim$0.14dex on our luminosity measurements.

\subsubsection{Departure from spherical symmetry}
Observations of RSG nebulae are often clumpy, rather than spherically symmetric \citep[e.g.][]{scicluna2015large,o2015alma,humphreys2007three}. We  investigated the effect of clumped winds by comparing our 1D models with those from MOCASSIN \citep{ercolano2003mocassin,ercolano2005dusty,ercolano2008x}, a code which solves the radiative transfer equation in 3D. We found that clumping has no effect up to a filling factor of 50. As long as the dust is optically thin there is no change in the output spectrum. 
\subsubsection{$T_{\rm in}$ and $\tau_V$}
Finally, DUSTY also allows inner temperature, T$_{\rm in}$, and the optical depth $\tau_V$ to be chosen. $T_{\rm in}$ defines the temperature of the inner dust shell (and hence it's position). The optical depth determines the dust shell mass. As these parameters are unconstrained, in this study we have allowed them to vary until the fit to the data is optimised. This fitting methodology is described in the next subsection. 

\subsection{Fitting methodology}
We first computed a grid of dust shell models spanning a range of inner temperatures and optical depths. For each model we then computed synthetic WISE and Spitzer photometry by convolving the model spectrum with their relevent filter profiles. This synthetic model photometry was compared to each stars mid-IR photometry from WISE, IRAC and MIPS. The grid spanned $\tau_{v}$ values of 0 - 1.3 with 50 grid points, and inner temperature values from 100K to 1200K in steps of 100K. By using ${\chi^2}$ minimisation we determined the best fitting model to the sample SED. 
\begin{equation}
\chi^2 =\sum \frac{ (O_{i}-E_{i})^2 }{\sigma_i^2}
\end{equation}
where O is the observed photometry, E is the model photometry, $\sigma$$^2$ is the error and i denotes the filter. In this case, the model photometry provides the "expected" data points. The best fitting model is that which produced the lowest  ${\chi^2}$. 

To account for systematic errors we applied a blanket error of 10\% to our observations. The errors on our best fitting model parameters were determined by models within our lowest $\chi^2$$\pm$10. This limit was chosen so that the stars with the lowest measured \mdot\, which were clearly consistent with non-detections, would have \mdot\ values consistent with 0 (or upper limits only).   

\section{Application to NGC2100}
In this study we apply this dust modeling to a sample of RSGs in a young star cluster. Such clusters can be assumed to be coeval, since any spread in the age of the stars will be small compared to the age of the cluster. Hence, we can assume that all stars currently in the RSG phase had the same initial mass to within a few tenths of a solar mass. Since the stars' masses are so similar, they will all follow almost the same path across the H-R diagram. Differences in luminosity are caused by those stars with slightly higher masses evolving along the mass-track at a slightly faster rate. It is for this reason that luminosity can be taken as a proxy for evolution.                                                                                                                                                                                 

\begin{figure*}
\caption{Finding chart for RSGs in NGC2100. The stars are numbered based on [5.6]-band magnitude.}
\centering
\label{fig:findingchart}
\includegraphics[width=\textwidth]{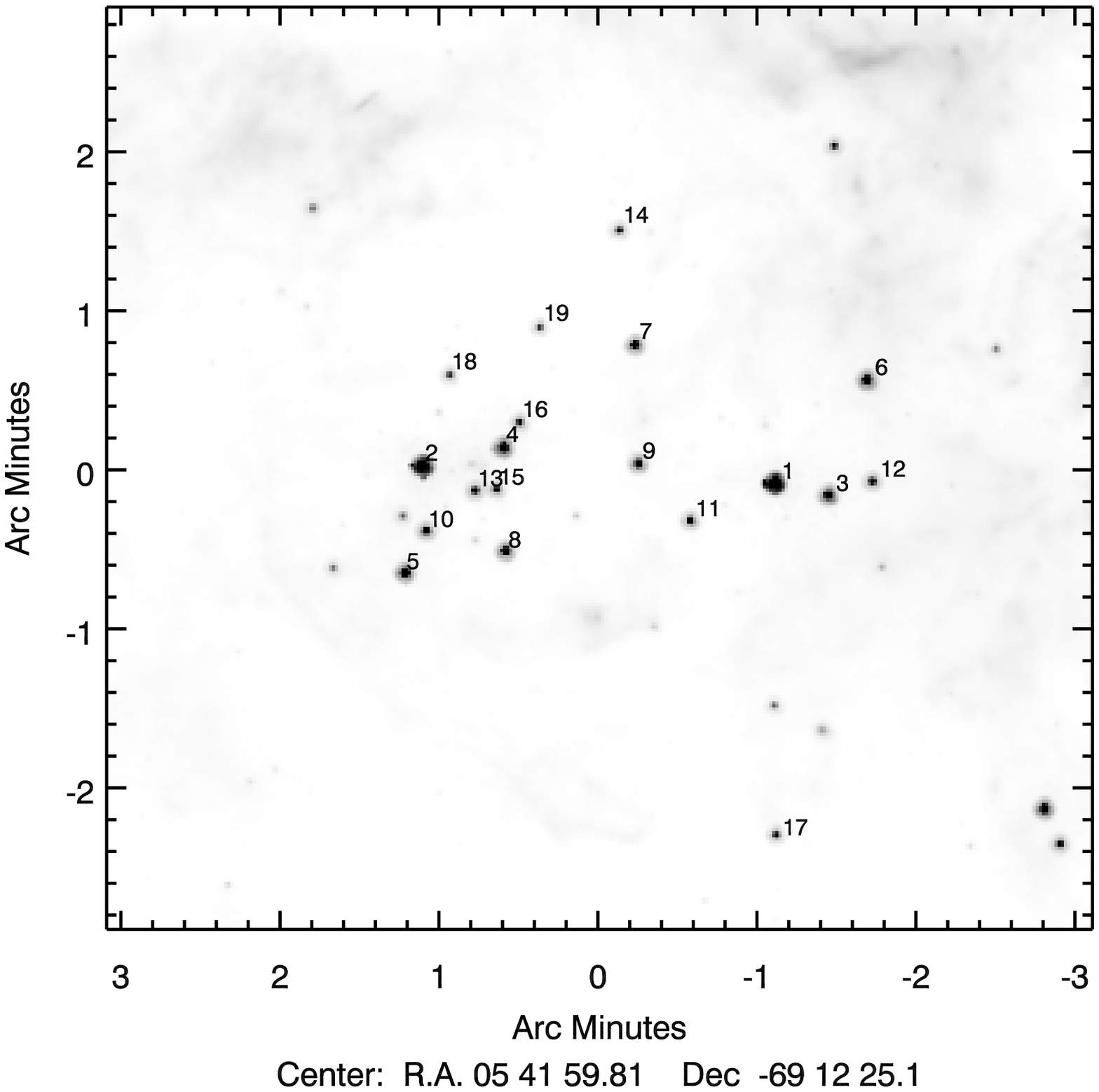}
\end{figure*}

The photometry used in this paper is taken from 2MASS, Spitzer and WISE \citep{Skrutskie2006two,werner2004spitzer,wright2010wide} and is listed in Table 2. A finding chart for NGC2100 is shown in Fig. \ref{fig:findingchart} in which the RSGs are numbered based on [5.6]-band magnitude. Star \#13 has been omitted from our analysis due to large disagreements between the MIPS and WISE photometry, as well as WISE and IRAC. 


\begin{table*}
\centering

\caption{Star designations and positions. Stars are numbered based on their [5.6]-band magnitude.}
\begin{tabular}{lccccc}
\hline\hline
Name & ID & RA ($\degr$) & DEC ($\degr$) & W61$^{a}$ & R74$^{b}$ \\
&&J2000 &J2000&&\\
\hline 
  1&     J054147.86-691205.9 & 85.44944763&-69.20166779 & 6-5 & D15 \\
  2&     J054211.56-691248.7 & 85.54819489&-69.21353149& 6-65 & B40 \\
  3&     J054144.00-691202.7 & 85.43335724&-69.20075989&8-67&...\\
  4&     J054206.77-691231.1 & 85.52821350&-69.20866394&...&A127 \\
  5&     J054209.98-691328.8 & 85.54161072&-69.22468567&6-51&C32 \\
  6&     J054144.47-691117.1 & 85.43533325&-69.18808746&8-70&...\\
  7&     J054200.74-691137.0 & 85.50312042&-69.19362640&6-30&C8 \\
  8&     J054203.90-691307.4 & 85.51628113&-69.21873474&6-34&B4 \\
  9&     J054157.44-691218.1 & 85.48937225&-69.20503235&6-24&C2 \\
 10&     J054209.66-691311.2 & 85.54025269&-69.21979523&6-54&B47 \\
 11&     J054152.51-691230.8 & 85.46879578&-69.20856476&6-12&D16 \\
 12&     J054141.50-691151.7 & 85.42295837&-69.19770813&8-63&... \\
 13&     J054207.48-691250.3 & 85.53116608&-69.21398163&6-48&... \\
 14&     J054204.78-691058.8 & 85.51993561&-69.18302917&6-44&... \\
 15&     J054206.13-691246.8 & 85.52555847&-69.21302032&6-46&... \\
 16&     J054206.36-691220.2 & 85.52650452&-69.20561218&6-45&B17 \\
 17&     J054138.59-691409.5 & 85.41079712&-69.23599243&8-58&... \\
 18&     J054212.20-691213.3 & 85.55084229&-69.20370483&6-69&B22 \\
 19&     J054207.45-691143.8 & 85.53106689&-69.19552612&6-51&C12 \\
 \multicolumn{6}{p{\columnwidth}}{$^{a}$ star designation from \cite{westerlund1961population}}\\
  \multicolumn{6}{p{\columnwidth}}{$^{b}$ star designation from \cite{robertson1974color}}\\

\end{tabular}
\end{table*}

\begin{table*}
\centering
\tiny
\caption{Observational data. All fluxes are in units of mJy.}
\label{my-label}
\begin{tabular}{lcccccccccccccccc}
\hline\hline
Name  & IRAC1 & IRAC2 & IRAC3 & IRAC4& MIPS1& 2MASS-J & 2MASS-H & 2MASS-Ks & WISE1& WISE2 & WISE3 & WISE4  \\
 &  (3.4  $\micron$) & (4.4  $\micron$) &(5.6  $\micron$) &(7.6$\micron$) &  (23.2  $\micron$)& & & & (3.4  $\micron$)& (4.6  $\micron$)& (11.6  $\micron$) &(22  $\micron$) \\
\hline
  1&176.0$\pm$0.2 & 163.00$\pm$ 0.03 & 150.0$\pm$0.05 & 141.00$\pm$0.04 & 140.00$\pm$0.063 & 244$\pm$ 4& 364$\pm$12 & 344$\pm$ 9 & 240$\pm$6 & 153$\pm$2.8 & 166$\pm$ 2.6 &  149$\pm$ 3.97  \\
  2& - &  95.10$\pm$ 0.02 &  115.0$\pm$0.05 & 120.00$\pm$0.03 & 113.00$\pm$0.063 & 237$\pm$ 5 & 359$\pm$13 & 324$\pm$ 5 & 173$\pm$  3 & 116$\pm$ 2.0 & 146$\pm$ 2.0 &  132$\pm$ 3.04  \\
  3&131.0$\pm$0.1& 72.70$\pm$ 0.02&  64.7$\pm$0.04 & 43.80$\pm$0.02 &    -& 232 $\pm$ 4 & 332$\pm$10 & 286$\pm$ 6 & 171 $\pm$  3 &  82$\pm$ 1.5 &  33$\pm$ 0.7 &   25$\pm$ 1.63 \\
  4& 86.6$\pm$0.1 & 65.20$\pm$ 0.02 &  60.7$\pm$0.04 & 56.50$\pm$0.02 &  33.20$\pm$ 0.063 & 161$\pm$ 3 & 220$\pm$ 4  & 196$\pm$ 4 & 120$\pm$  3 & 65$\pm$ 1.5 &  51$\pm$ 0.8 &  36$\pm$ 1.59 \\
  5&101.0$\pm$0.1 &  64.10$\pm$ 0.02 &  56.9$\pm$0.03 & 49.00$\pm$0.02 & 26.10$\pm$0.064 & 154$\pm$ 3 & 220$\pm$ 4 & 194$\pm$ 4 & 145$\pm$  2 &  66$\pm$ 1.2 &  39$\pm$ 0.7 &  30$\pm$ 1.30 \\
  6&110.0$\pm$0.1& 65.10$\pm$ 0.02 &  56.6$\pm$0.03 & 46.00$\pm$0.02 & 27.50$\pm$0.065 & 186$\pm$ 3 & 270$\pm$ 6 & 240$\pm$ 4& 137$\pm$  2 &  70$\pm$ 1.3 &  37$\pm$ 1.0 &  35$\pm$ 3.08 \\
  7& 92.6$\pm$0.1 &  62.30$\pm$ 0.02 &   53.7$\pm$0.03 & 38.10$\pm$0.02 &     -&173$\pm$ 3&249$\pm$ 5&220$\pm$ 4& 125$\pm$  2 &  60$\pm$ 1.1 & 27$\pm$ 0.6 &   12$\pm$ 1.23 \\
  8& 93.5$\pm$0.1& 58.10$\pm$ 0.02 &  50.9$\pm$0.03 & 42.00$\pm$0.02 &  19.20$\pm$0.063 & 183$\pm$ 3 & 254$\pm$ 5 & 209$\pm$ 4 & 113$\pm$  2 &  56$\pm$ 0.9 & 30$\pm$ 0.5 &   14$\pm$ 1.63 \\
  9& 93.6$\pm$0.1 & 58.30$\pm$ 0.02 &  48.4$\pm$0.03 &  36.30$\pm$0.02 &     -& 187$\pm$ 3 & 244$\pm$ 5 & 209$\pm$ 4 & 113$\pm$  2 &  56$\pm$ 1.0 &  22$\pm$ 0.6 &    2$\pm$ 0.77 \\
 10& 80.7$\pm$0.1& 50.70$\pm$ 0.02 &  41.3$\pm$0.03 & 26.00$\pm$0.02 &   -&161$\pm$ 3 & 223$\pm$ 4 & 190$\pm$ 4& 105$\pm$  2 & 51$\pm$ 1.0 & 15$\pm$ 0.3 &  11$\pm$ 1.14 \\
 11& 68.1$\pm$0.1 &  43.30$\pm$ 0.01 &   34.7$\pm$0.03 & 23.70$\pm$0.02 &    -& 108$\pm$ 2 & 156$\pm$ 3 & 143$\pm$ 2 &  84$\pm$  1 &  41$\pm$ 0.8 &  10$\pm$ 0.5 &    4$\pm$ 0.87 \\
 12& 63.7$\pm$0.1 &  41.80$\pm$ 0.01 &   32.8$\pm$0.03 &  23.00$\pm$0.02 &     - & 125$\pm$ 2 & 181$\pm$ 3 & 156$\pm$ 3 &  83$\pm$  1 & 39$\pm$ 0.7 &  13$\pm$ 0.5 &    3$\pm$ 1.25 \\
 13& 75.2$\pm$0.1 & - &   32.4$\pm$0.03 &  21.50$\pm$0.02 &     - & 131$\pm$ 3 & 178$\pm$ 4 & 160$\pm $3 & 140$\pm$  2 &  73$\pm$ 1.3 & 20$\pm$ 0.4 &   21$\pm$ 1.07 \\
 14& 65.7$\pm$0.1 &  37.50$\pm$ 0.01 &   29.2$\pm$0.03 &  16.60$\pm$0.02 &     - &117$\pm$ 2 & 171$\pm$ 3 & 142$\pm$ 2 &  75$\pm$  1& 35$\pm$ 0.6 &   6$\pm$ 0.4 &   - \\
 15& 62.9$\pm$0.1 & 36.30$\pm$ 0.01 &   28.7$\pm$0.02 &  17.80$\pm$0.02 &     - & 120$\pm$ 3 & 166$\pm$ 5 & 137$\pm$ 3 &106$\pm$  9 & 44$\pm$ 3.7 &   4$\pm$ 1.2 &   - \\
 16& 62.0$\pm$0.1 &  37.00$\pm$ 0.01 &   28.4$\pm$0.02 &  17.80$\pm$ 0.02 &     -& 112$\pm$ 2 & 162$\pm$ 3 & 142$\pm$ 2 &  94$\pm$  1 & 49$\pm$ 0.9 & 19$\pm$ 0.4 &   14$\pm$ 1.03 \\
 17& 59.7$\pm$0.1 &  35.70$\pm$ 0.01 &  27.1$\pm$0.02 &  16.80$\pm$0.02 &     -&113$\pm$ 2 & 155$\pm$ 3 & 135$\pm$ 2 &  73$\pm$  1 &  34$\pm$ 0.6 &  9$\pm$ 0.3 &    3$\pm$ 1.08 \\
 18& 51.4$\pm$0.1 &  31.50$\pm$ 0.01 &   24.0$\pm$0.02 &  14.40$\pm$0.01 &    - & 105$\pm$ 2 & 142$\pm$ 3 & 121$\pm$ 2 &  64$\pm$  1 &  30$\pm$ 0.6 &   5$\pm$ 0.2 &    - \\
 19& 53.8$\pm$0.1 &  30.40$\pm$ 0.01 &  22.6$\pm$0.02 &    - &     - & 101$\pm$ 2 & 144$\pm$ 2 & 119$\pm$ 2 &  61$\pm$  1 &  28$\pm$ 0.5 &   3$\pm$ 0.2 &    - \\

\end{tabular}
\end{table*}

\begin{figure}
  \caption{Colour-magnitude plot using J-K$_{\rm s}$ vs. K$_{\rm s}$ to locate RSGs in NGC 2100. This plot also shows a14Myr PARSEC isochrone \citep{tang2014new,chen2015parsec} at LMC metallicity (non-rotating). Isochrones have been adjusted for the distance to the LMC and a foreground extinction of $A_V$=0.5. The extinction noted in the legend is {\it in addition }to the foreground extinction already known to be present in the LMC \citep[][]{niederhofer2015no}. }
  \centering
  \label{fig:colcol}
    \includegraphics[width=\columnwidth]{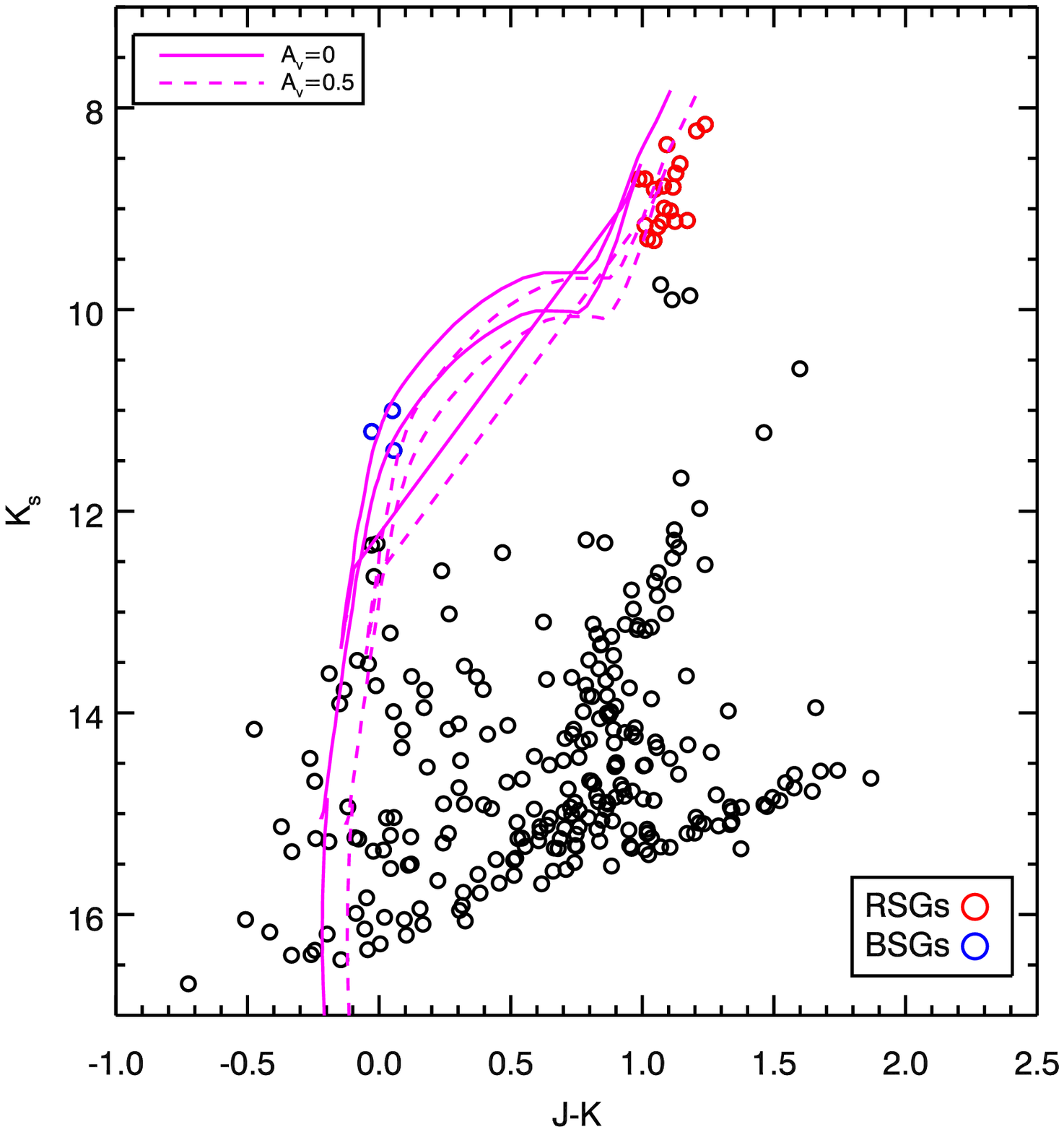}
\end{figure}

The RGSs in NGC2100 can be seen as a clump of stars in CMD space with a $K_S$-band magnitude less than 9.49 within a 2 arcminute radius of the cluster centre. This identified 19 candidate RSGs. By plotting J-K vs. K it was possible to locate RSGs in the data as a clump of stars clearly separated from the field stars. This is shown in Fig. \ref{fig:colcol}. The red circles indicate RSGs. 

\begin{figure}
  \caption{Colour magnitude plot of RSGs in the cluster to show increasing dust mass with age. [5.6]-band magnitude is used as an indicator of $L_{\rm bol}$ and the [8-12] colour is used as a measure of dust shell mass. The [8-12] colour is useful as is includes the mid-IR excess and the excess caused by the broad silicate feature.}
  \centering
  \label{fig:8min12}
    \includegraphics[width=\columnwidth]{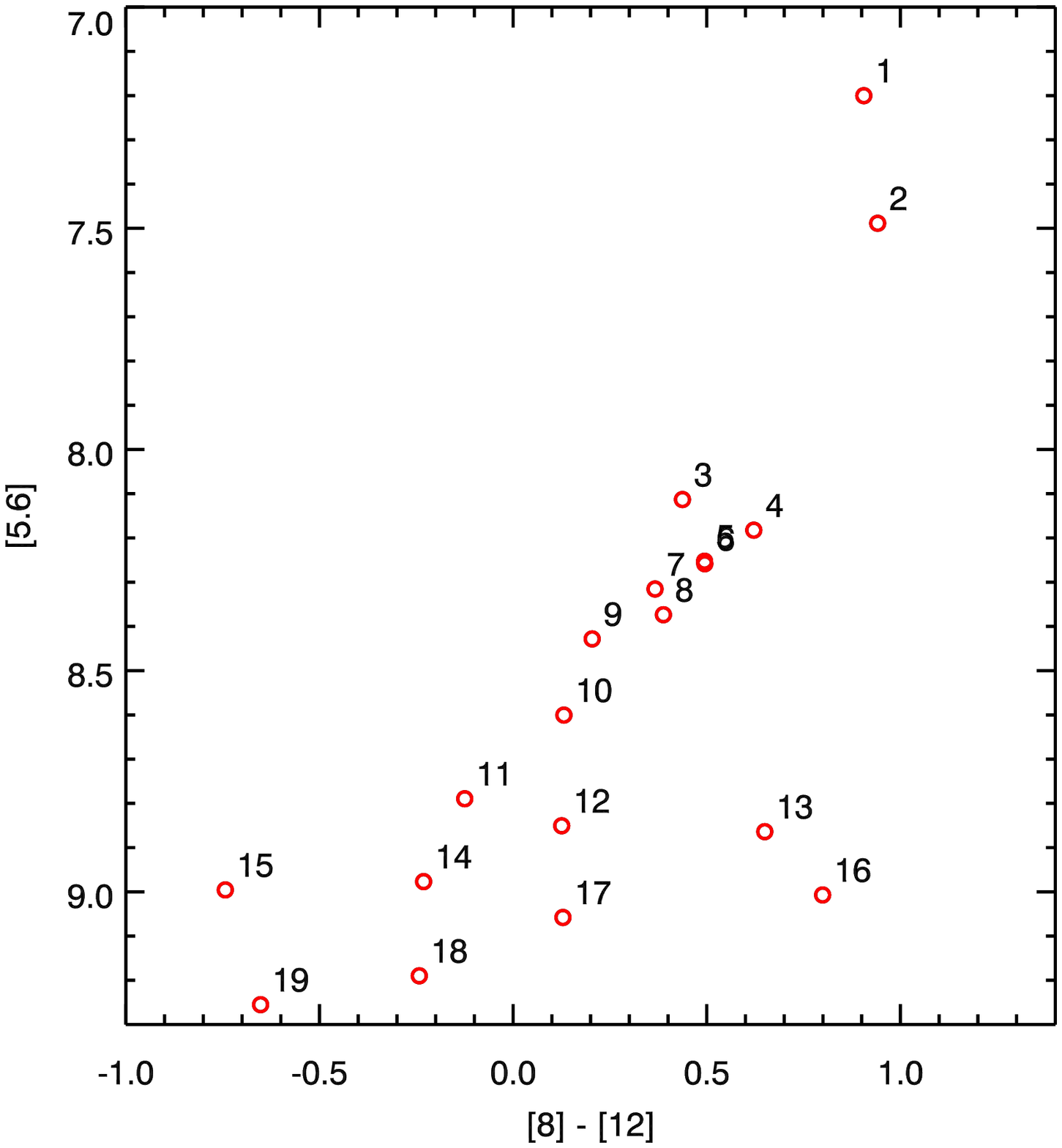}
\end{figure}

From the photometry alone it was possible to see evidence of \mdot\ evolving with evolution of the RSG. This qualitative evidence is shown in the [8-12] vs. [5.6] CMD, Fig. \ref{fig:8min12}. The 5.6-band magnitude can be used as a measure of luminosity as the bolometric correction at this wavelength is largely insensitive to the RSGs temperatures, whilst also being too short a wavelength to be significantly affected by emission from circumstellar dust. The [8-12] colour can be used as a measure of dust shell mass as it measures the excess caused by the broad silicate feature at 10$\micron$. It can be seen from Fig. \ref{fig:8min12} that more luminous (and therefore, more evolved) RSGs have a larger amount of dust surrounding them (shown by the increasing colour, meaning they appear more reddened), suggesting dust mass increases with age.

Below we discuss our modeling results and compare them to mass-loss rate prescriptions frequently used by stellar evolution groups.

\subsection{Modeling results}
\begin{figure*}
  \caption{ \textit{Left panel:}  Model plot for the star with the highest \mdot\ value in NGC 2100 including all contributions to spectrum. The "error models" are the models that fit within the minimum $\chi^2$+10 limit. The stars are numbered based on 5.6-band magnitude (\#1 being the star with the highest [5.6]-band magnitude). The silicate bump at 10$\micron$ is clearly visible on the spectra suggesting a large amount of circumstellar material. \textit{Right panel:} Contour plot showing the degeneracy between $\chi^2$ values and best fitting \mdot\ values in units of 10$^{-6}$ M$_\odot$ yr$^{-1}$. The green lines show the best fit \mdot\ and upper and lower \mdot\ isocontours. It can be seen that while there is some degeneracy between inner dust temperature and optical depth the value of \mdot\ is independent of this. }
  \centering
  \label{fig:allcont}
     \includegraphics[height=7cm,bb=60 0 850 566,clip]{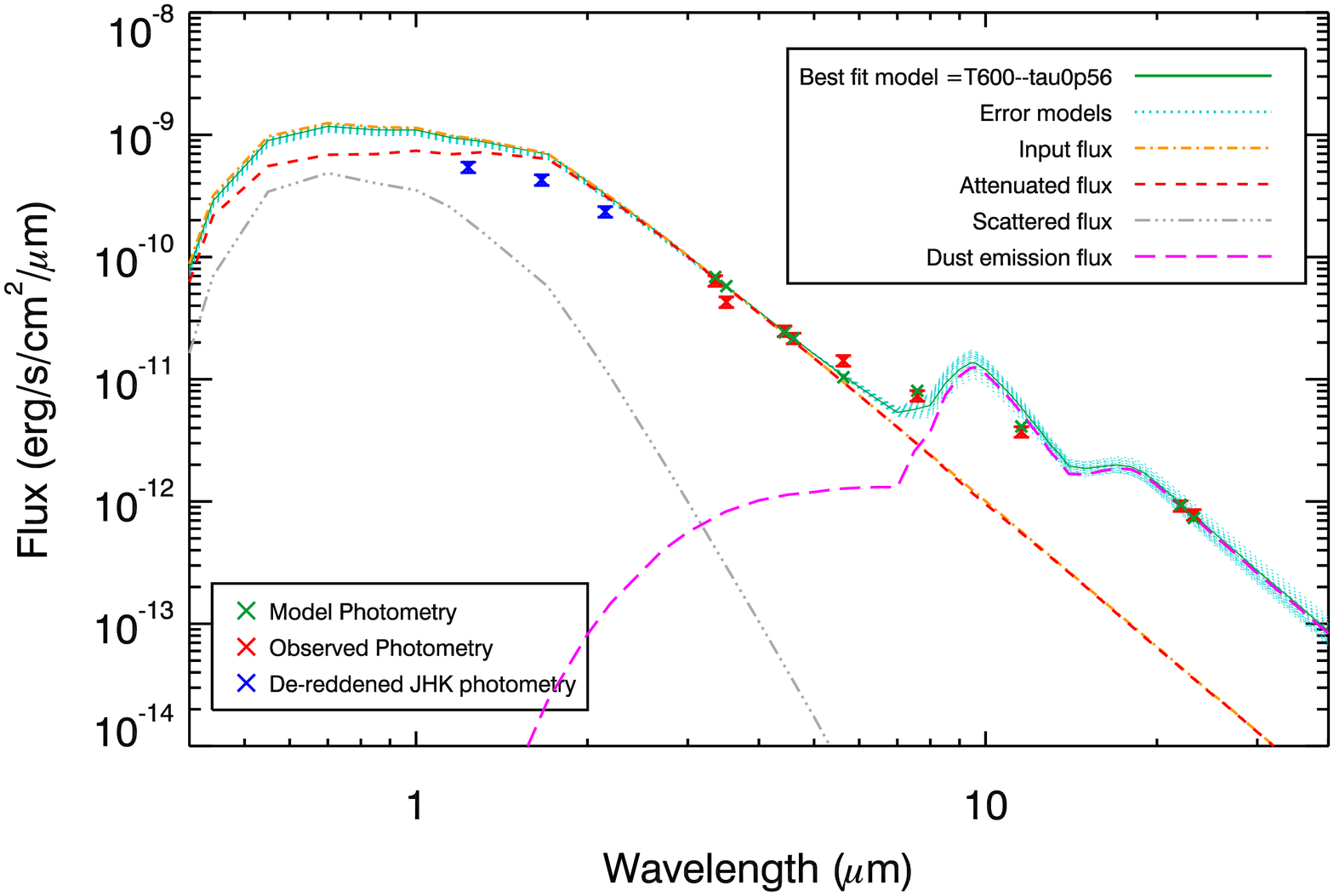}
     \includegraphics[height=7cm]{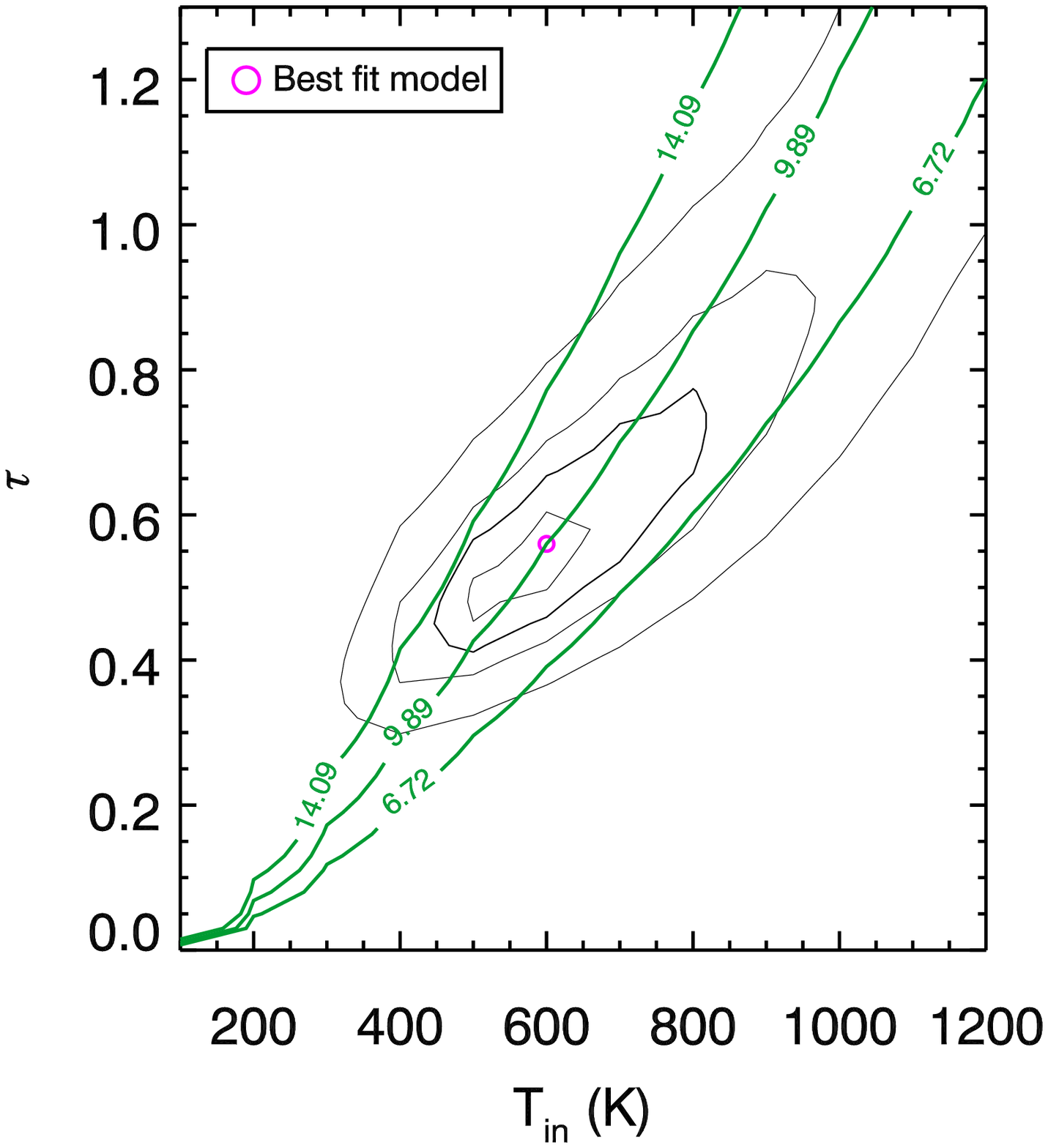}
\end{figure*}

\begin{figure*}
  \caption{Same as Fig. \ref{fig:allcont} for star \#8, which has an intermediate \mdot\ value. It can be seen in the model plot (left) that it is possible to fit both the near-IR and mid-IR photometry.  \mdot\ values are in units of 10$^{-6}$ M$_\odot$ yr$^{-1}$.}
  \centering
  \label{fig:medmdot}
    \includegraphics[height=7cm,bb=60 0 850 566,clip]{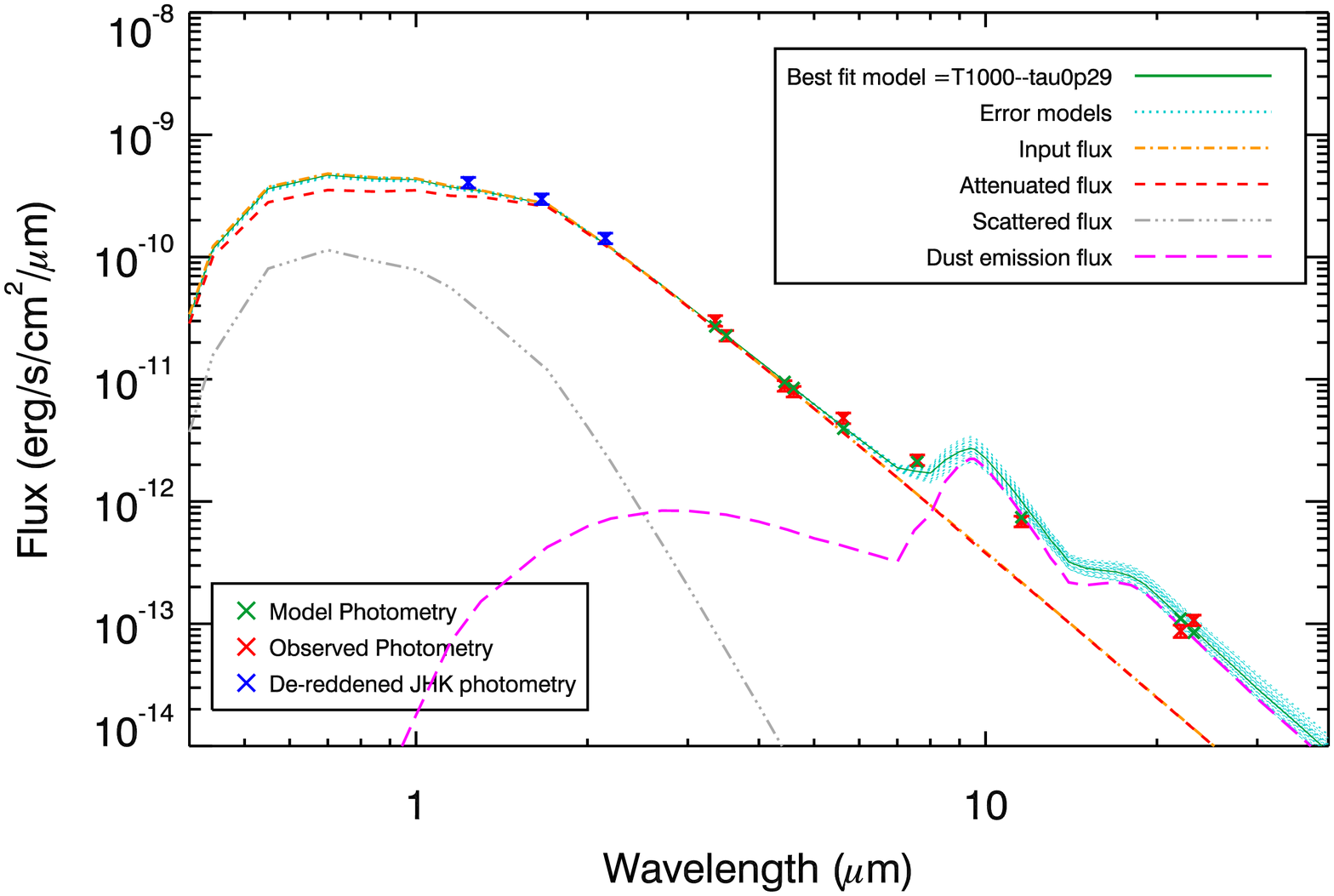}
    \includegraphics[height=7cm]{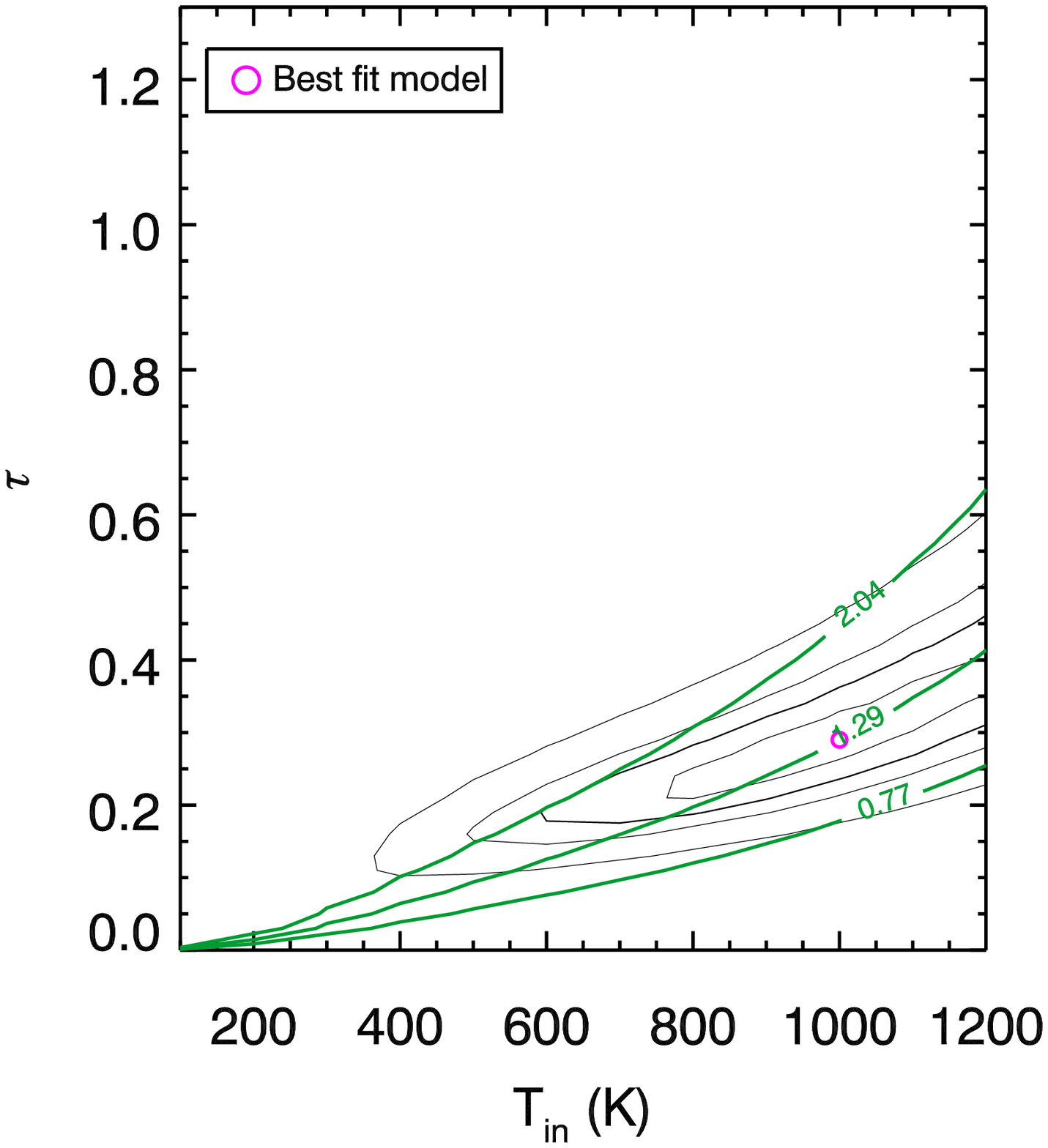}
    
\end{figure*}

\begin{figure*}
  \caption{Same as Fig. \ref{fig:allcont} for star \#12, which has a low \mdot\ value. It can be seen in this plot that it is possible to fit both the near-IR photometry and mid-IR photometry. \mdot\ values are in units of 10$^{-6}$ M$_\odot$ yr$^{-1}$.}
  \centering
  \label{fig:lowmdot}
    \includegraphics[height=7cm,bb=60 0 850 566,clip]{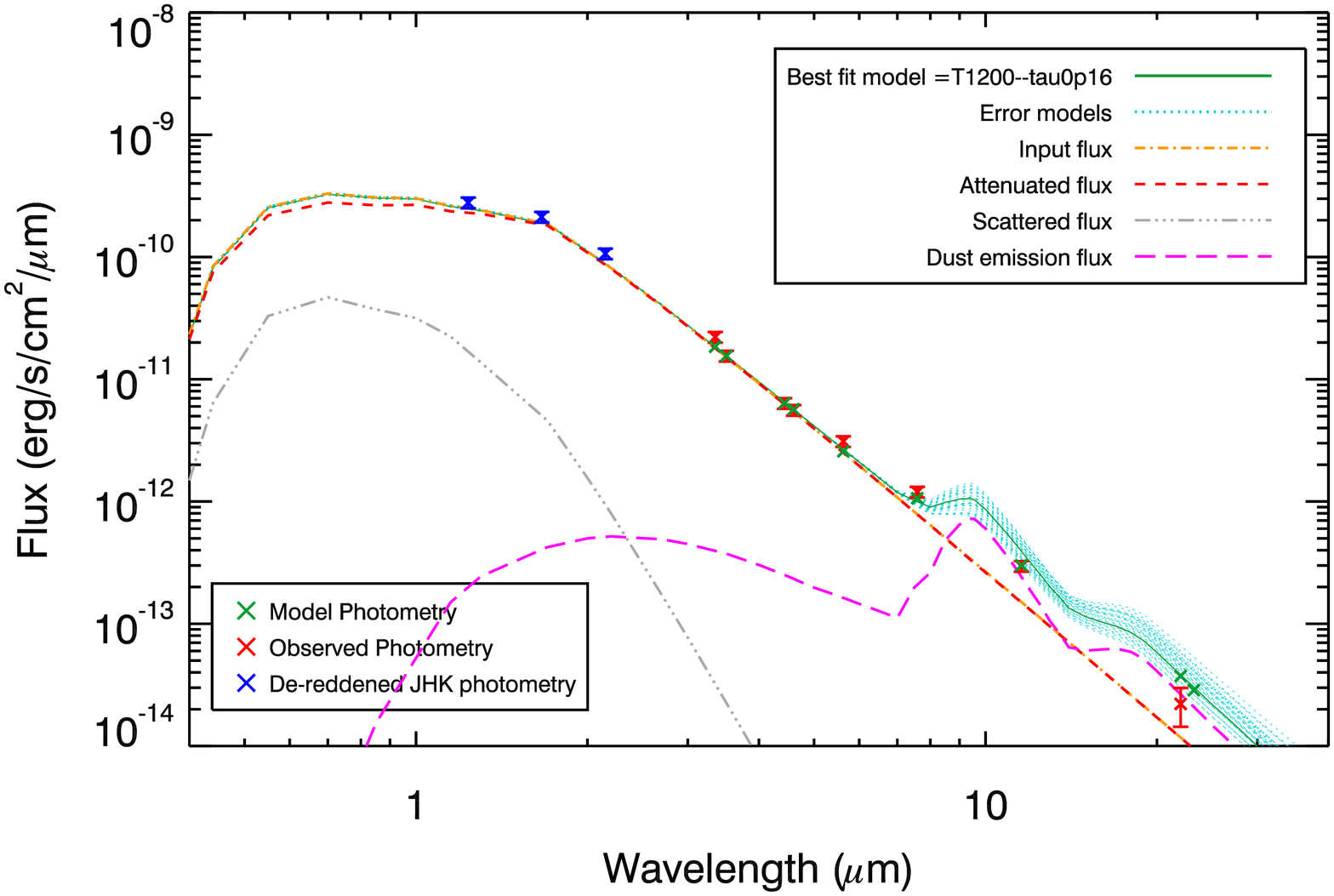}
     \includegraphics[height=7cm]{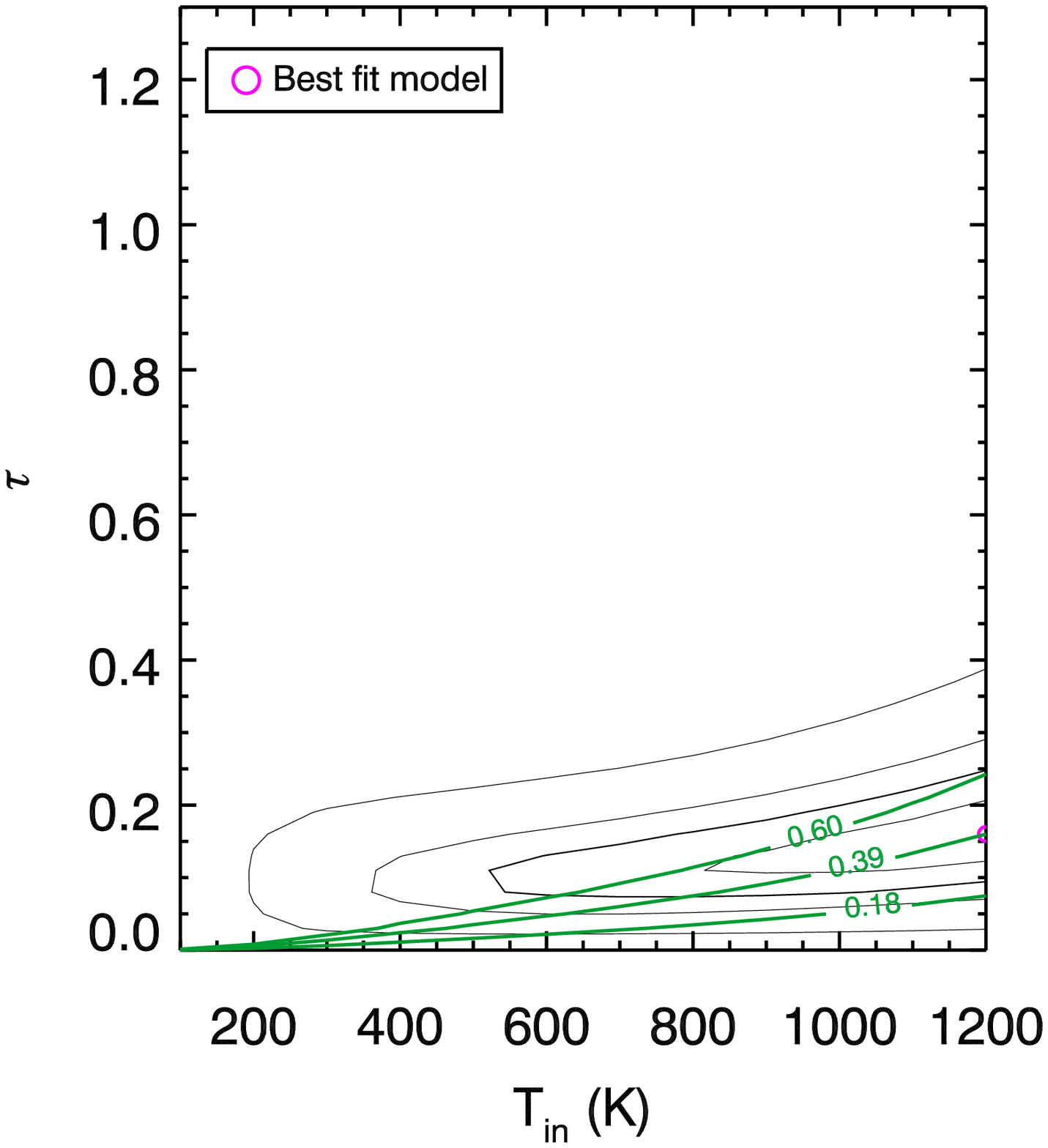}
\end{figure*}

We ran our fitting procedure for the 19 RSGs located in NGC2100, our results are shown in Table 2. Figures \ref{fig:allcont}-\ref{fig:lowmdot} shows an example model fit with observed photometry for star \#1. The plot shows our best fit model spectra (green line), the models within our error range (blue dotted lines) and the various contributions to the flux, including scattered flux, dust emission and attenuated flux. It also shows the photometric data (red crosses) and model photometry (green crosses). The 10$\micron$ silicate bump can be clearly seen due to dust emission (pink dashed line). The plot also shows the significant effect scattering within the dust shell (grey dotted/dashed line), contributing to a large proportion of the optical output spectrum.

The fitting procedure did not include the JHK photometry bands as these bands are strongly affected by extinction, but when over-plotting this photometry (once de-reddened) it was found to be in good agreement with the model spectrum for all stars except \#1 and \#2, for which the model over-predicts the near-IR flux. This deficit in the observed NIR flux is not present for the other RSGs in our sample. Figures \ref{fig:medmdot} and  \ref{fig:lowmdot} show the model fits for stars \#8 and \#12 respectively, representative of medium and low \mdot\ values. 

We attempted to explain the missing near-IR (NIR) flux in stars \#1 and \#2 by adapting the fitting procedure to include JHK photometry and to include a lower \teff\ SED. This gave us a better fit to the near-IR photometry but at the expense of a poorer fit to the 3-8$\micron$ region, where the model now underpredicted the flux. We considered whether this fit could be improved by dust emission. To achieve this, it would require either unphysically high dust temperatures above the sublimation temperature for silicate dust, or it would require an increase in dust mass of a factor of 100. This would lead to significantly poorer fits to the mid-IR photometry and can therefore be ruled out. There was only a small effect on the best fit \mdot\ found (less than 10\%). The only change to our results from making these adjustments was that the $L_{\rm bol}$ was reduced for stars \#1 and \#2 by approximately 0.3 dex. We discuss these results further in Section 4.1. 

In Fig. \ref{fig:allcont} we show a contour plot illustrating the degeneracy between our two free parameters, $T_{\rm in}$ and $\tau_{\rm v}$, with \mdot\ contours for the best fit \mdot\ and upper and lower \mdot\ contours overplotted. It can be seen that the lines of equal \mdot\ run approximately parallel to the $\chi^2$ contours. This means that despite the degeneracy between $\tau_V$ and $T_{\rm in}$ the value of \mdot\ is well constrained and robust to where we place the inner dust rim. 

Fit results for all stars modelled are shown in Table 2. We find a varying $T_{\rm in}$ value for the stars in the sample, rather than a constant value at the dust sublimation temperature of 1200K. For each of the stars a best fit value of $\tau$ and $T_{\rm in}$ is found. Lower $T_{\rm in}$ values have also been found in other studies \citep[c.f.][]{groenewegen2009luminosities}. When compared to the stars' calculated luminosities, it can be seen that lower luminosity stars have a greater spread in $T_{\rm in}$ values, while higher \mdot\ stars have $T_{\rm in}$ values that are more constrained. We find that all stars in our sample are consistent with $T_{\rm in}$ $\sim$ 600K. $L_{\rm bol}$ is found by integrating under the model spectra with errors on $L_{\rm bol}$ dominated by the uncertainty in \teff. The value of $A_V$ is found from the ratio of input and output fluxes at 0.55$\micron$ and is intrinsic to the dust shell. For stars numbered 15, 18 and 19 the value of \mdot\ is so low it can be considered as a non-detection, leaving $T_{\rm in}$ unconstrained.

 A positive correlation between \mdot\ and luminosity is illustrated in Fig. \ref{fig:mdotvsL}, implying that \mdot\ increases by a factor of 40 during the RSG phase, which according to model predictions should last approximately 10$^6$ years for stars with initial masses of 15M$_\odot$ \citep{georgy2013populations}, see Section 3.2. This plot also shows some mass loss rate prescriptions for comparison (assuming a \teff of 4000K); \cite{de1988mass}(hereafter dJ88) , \cite{reimers1975circumstellar,kudritzki1978absolute}, \cite{van2005empirical},\cite{nieuwenhuijzen1990parametrization}(hereafter NJ90) and \cite{feast1992ch}. See Section 3.1.1 for further discussion of the \mdot\ prescriptions. We find our results are best fit by dJ88, van Loon and Reimer's prescriptions, with dJ88 providing a better fit for the more evolved stars (where the mass loss mechanism is stronger). 
 
Our stars form a tight correlation, whereas previous studies of \mdot\ with $L_{\rm bol}$ (e.g. \cite{van2005empirical}) have shown a large spread in results. This could be due to previous studies looking at field stars, whereas our study has looked at RSGs in a coeval cluster. As for the three stars with negligible \mdot\ values, it is possible that no appreciable amount of dust has formed around these RSGs yet meaning the dust driven wind hasn't taken effect. We considered the possibility that these stars were foreground stars but after checking their v$_{rad}$ values \citep{patrick2016chemistry} we find they are all consistent with being within the cluster.

 \subsubsection{Mass loss rate prescriptions}
\begin{figure*}
\centering
  \caption{Plot showing \mdot\ versus $L_{\rm bol}$. A positive correlation can be seen suggesting \mdot\ increases with evolution. This is compared to some mass loss rate prescriptions. The downward arrows show for which stars we only have upper limits on \mdot.}
  \centering
  \label{fig:mdotvsL}
    \includegraphics[width=\textwidth]{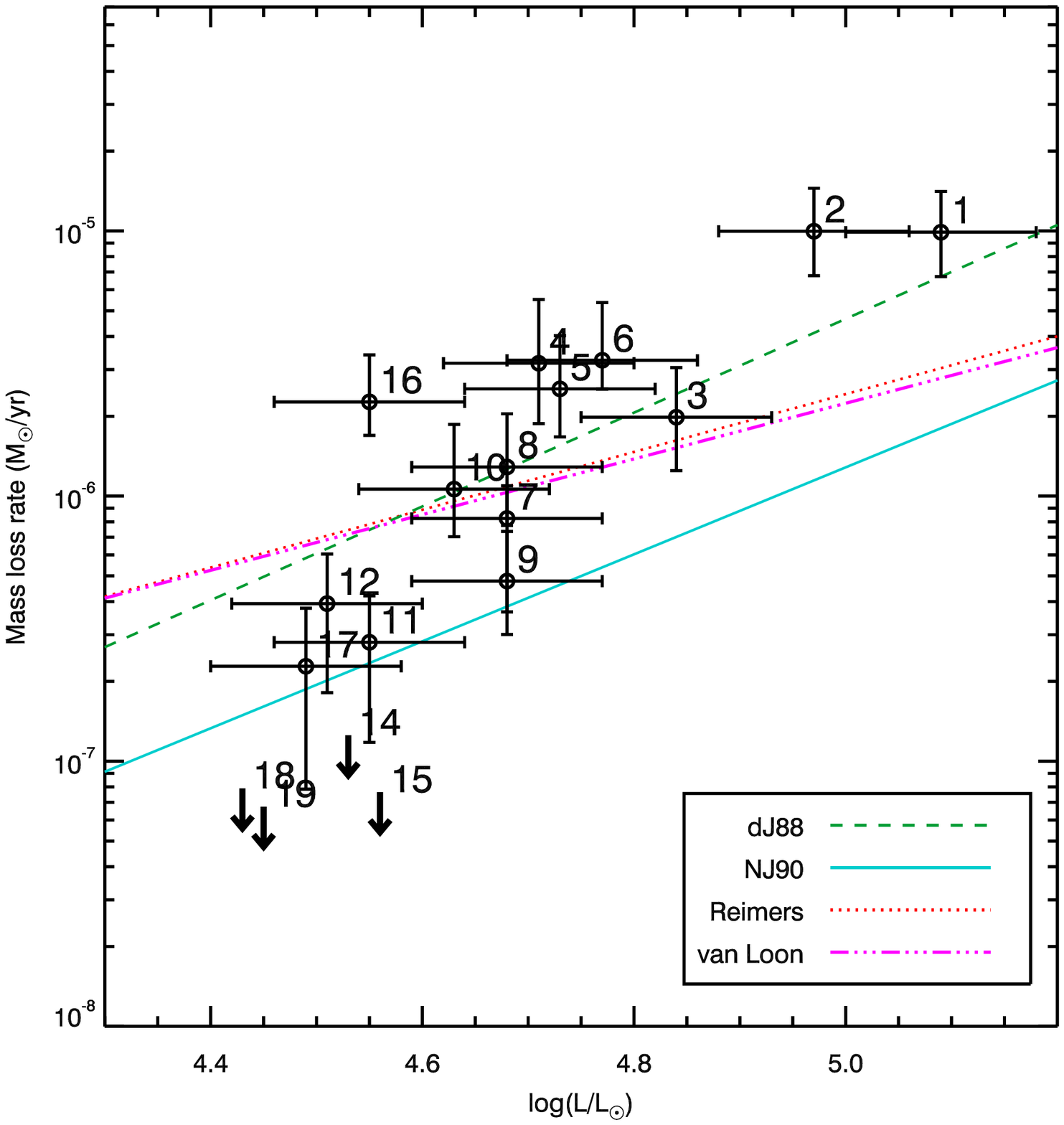}
\end{figure*}

 Each \mdot\ prescription was calculated using different methods. The \teff\ was set to 3900K for all prescriptions shown in Fig. \ref{fig:mdotvsL}.
 
 The empirical formula for dJ88 was derived by comparing \mdot\ values found from 6 different methods from literature for 271 stars of spectral types O through M. Determination of \mdot\ for M type stars included the modeling of optical metallic absorption lines of nearby RSGs \citep[under the assumption the lines form in the wind][]{sanner1976mass} and using mid-IR photometry and hydrodynamics equations to find $v_{\infty}$ \citep{gehrz1971mass}. This relation is a two parameter fit on of $L_{\rm bol}$ and \teff. \cite{de1988mass} found that each method found the same \mdot\ value to within the error limits no matter the star's position on the HR-diagram. The NJ90 prescription \citep{nieuwenhuijzen1990parametrization} is a second formulation of the dJ88 formula, including stellar mass. Due to the narrow mass range for RSGs (8-25M$_\odot$) and the very weak dependence on M it has very little effect on the \mdot\ found from this formulation.

 Reimer's law \citep{reimers1975circumstellar,kudritzki1978absolute} is a semi-empirical formula derived by measurements of circumstellar absorption lines for companions in binary systems. This has been repeated for 3 such systems only but provides an accurate measurement of \mdot.  The formula depends on surface gravity, $g$, but can be expressed in terms of $R$,$L$ and $M$ (in solar units) as shown by \cite{mauron2011mass}.
 
 Van Loon's prescription is an empirical formula based on optical observations of dust enshrouded RSGs and Asymptotic Giant Branch (AGB) stars within the LMC, where 
 \mdot\ was derived by modeling the mid-IR SED using DUSTY. \cite{van2005empirical} assumed a constant grain size of 0.1$\micron$, but state that this value was varied for some of the stars to improve fits. $T_{\rm in}$ values were first assumed to be between 1000 and 1200K, but again the author states for some stars in the sample this was reduced to improve the fit of the data and the DUSTY model.

The most widely used \mdot\ prescription, dJ88, provides the best fit to our observations for the more evolved stars. The van Loon prescription also agrees quite well, even though one might expect that this study's focus on dust-enshrouded stars would be biased towards higher \mdot\ stars. All of the prescriptions over predict the \mdot\ for the lowest luminosity stars. Though, this may be due to the fact that dust in these stars has yet to form (and hence $r_{gd}$ > 500 for these).

\begin{table*}
\caption{Results for stars in NGC 2100. Stars are numbered with \#1 having the highest [5.6]-band magnitude and \#19 having the lowest. Luminosities quoted are in units of log($L_{\rm bol}$/L$_\odot$). $A_V$ is the extinction intrinsic to the dust shell.}.
\centering

\begin{tabular}{lcccccc}

\hline\hline
Star & $T_{\rm in}$ (K) & $\tau_V$ &  \mdot\ (10$^{-6}$M$_\odot$ yr$^{-1}$) &  $L_{\rm bol}$   &$A_V$  \\ [0.5ex] 
\hline
 1&$ 600^{+ 200}_{- 100}$&$0.56^{+0.21}_{-0.14}$&$ 9.89^{+ 4.20}_{- 3.17}$&$ 5.09\pm 0.09$&$0.09^{+0.04}_{-0.02}$ \\
 2&$ 600^{+ 200}_{- 100}$&$0.64^{+0.26}_{-0.16}$&$ 9.97^{+ 4.52}_{- 3.19}$&$ 4.97\pm 0.09$&$0.10^{+0.05}_{-0.03}$ \\
 3&$ 600^{+ 400}_{- 200}$&$0.16^{+0.08}_{-0.05}$&$ 1.98^{+ 1.07}_{- 0.74}$&$ 4.84\pm 0.09$&$0.02^{+0.01}_{-0.01}$ \\
 4&$ 800^{+ 400}_{- 200}$&$0.45^{+0.32}_{-0.16}$&$ 3.17^{+ 2.34}_{- 1.29}$&$ 4.71\pm 0.09$&$0.07^{+0.06}_{-0.03}$ \\
 5&$ 700^{+ 300}_{- 200}$&$0.29^{+0.16}_{-0.08}$&$ 2.54^{+ 1.49}_{- 0.86}$&$ 4.73\pm 0.09$&$0.04^{+0.03}_{-0.01}$ \\
 6&$ 500^{+ 300}_{- 100}$&$0.21^{+0.13}_{-0.02}$&$ 3.25^{+ 2.12}_{- 0.72}$&$ 4.77\pm 0.09$&$0.03^{+0.02}_{-0.00}$ \\
 7&$1200^{+   0}_{- 500}$&$0.27^{+0.07}_{-0.14}$&$ 0.82^{+ 0.27}_{- 0.46}$&$ 4.68\pm 0.09$&$0.04^{+0.01}_{-0.02}$ \\
 8&$1000^{+ 200}_{- 400}$&$0.29^{+0.16}_{-0.10}$&$ 1.29^{+ 0.76}_{- 0.51}$&$ 4.68\pm 0.09$&$0.04^{+0.03}_{-0.01}$ \\
 9&$1200^{+   0}_{- 400}$&$0.16^{+0.08}_{-0.05}$&$ 0.48^{+ 0.26}_{- 0.18}$&$ 4.68\pm 0.09$&$0.02^{+0.01}_{-0.01}$ \\
10&$ 600^{+ 500}_{- 200}$&$0.11^{+0.08}_{-0.03}$&$ 1.06^{+ 0.80}_{- 0.36}$&$ 4.63\pm 0.09$&$0.01^{+0.01}_{-0.00}$ \\
11&$1200^{+   0}_{- 700}$&$0.11^{+0.05}_{-0.06}$&$ 0.28^{+ 0.14}_{- 0.16}$&$ 4.55\pm 0.09$&$0.01^{+0.01}_{-0.01}$ \\
12&$1200^{+   0}_{- 600}$&$0.16^{+0.08}_{-0.08}$&$ 0.39^{+ 0.21}_{- 0.21}$&$ 4.51\pm 0.09$&$0.02^{+0.01}_{-0.01}$ \\
14&$-$&$<0.05$&$<0.12$&$ 4.53\pm 0.10$&$-$ \\
15&$-$&$<0.03$&$ <0.08$&$ 4.56\pm 0.09$&$-$ \\
16&$ 400^{+ 300}_{- 100}$&$0.13^{+0.06}_{-0.02}$&$ 2.27^{+ 1.14}_{- 0.57}$&$ 4.55\pm 0.09$&$0.020^{+0.010}_{-0.000}$ \\
17&$1100^{+ 100}_{- 700}$&$0.08^{+0.05}_{-0.05}$&$ 0.23^{+ 0.15}_{- 0.15}$&$ 4.49\pm 0.09$&$0.010^{+0.010}_{-0.010}$ \\
18&$-$&$<0.03$&$<0.08$&$ 4.43\pm 0.09$&$-$ \\
19&$-$&$<0.03$&$<0.07$&$ 4.45\pm 0.11$&$-$ \\
\hline

\end{tabular}
\end{table*}

\subsection{Cluster age and initial masses}
\begin{figure}
\centering
\caption{Plot showing $M_{\rm initial}$ vs luminosity for various mass tracks. The plot shows the upper and lower luminosity values at each $M_{\rm initial}$ for STARS \citep{eldridge2009spectral} (pink solid lines) at $Z \sim 0.008$, \citep{brott2011rotating} non-rotating models at LMC metallicity (green dashed line), and Geneva rotating (red dotted line) and non-rotating (blue dotted line) \citep{georgy2013populations} at LMC metallicity. The Geneva models do not currently cover masses greater than 15$M_\odot$ at this metallicity. The grey shaded region shows the upper and lower luminosities derived for the stars in our sample. Using our lowest measured $L_{\rm bol}$ of $\sim 4.5 L_\odot$ as a constraint we find $M_{\rm initial}$ of $\sim 14 M_{\odot} - 17 M_{\odot}$ from the evolutionary models.}
\centering
\label{fig:masstrack}
\includegraphics[width=\columnwidth]{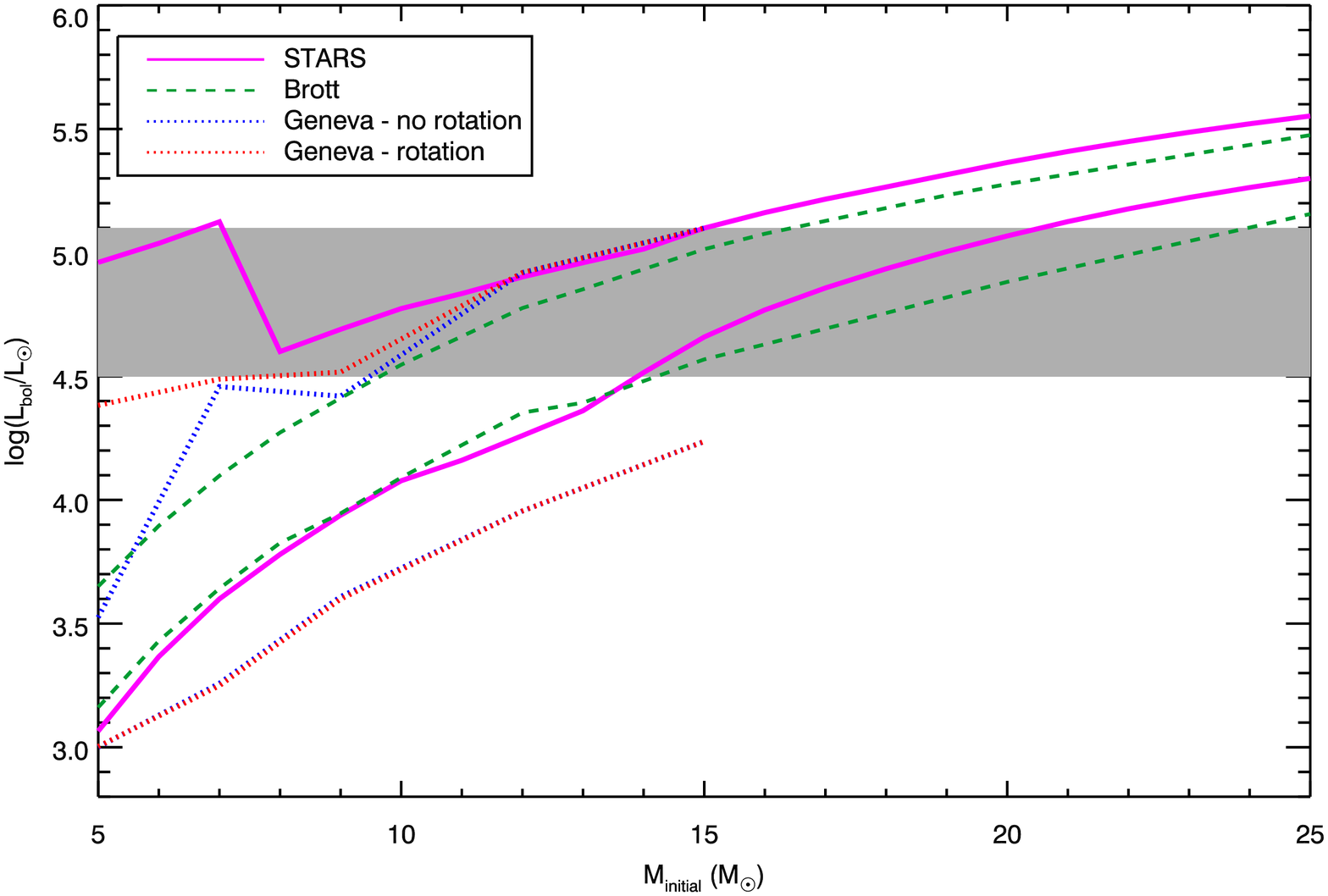}
\end{figure}

It was necessary to know the initial masses of the stars in our cluster. By comparing various stellar evolutionary models, such as \citep{brott2011rotating}, STARS \citep{eldridge2009spectral} and Geneva \citep{georgy2013populations}, and our lowest measured $L_{\rm bol}$ of $\sim$4.5L$_\odot$ as a constraint, see Fig. \ref{fig:masstrack} we derived an initial mass for the stars in the sample . It should be noted that \cite{brott2011rotating}'s mass tracks are not evolved to the end of Helium burning, but since we are only interested in the initial mass of the cluster stars this does not affect our conclusions. The current Geneva models at LMC metallicity are currently only available for masses up to 15M$_\odot$, but seem to imply a mass greater than 14M$_\odot$. We conclude that an $M_{\rm initial}$ of $\sim$14M$_{\odot}$ - 17M$_\odot$ seems most likely.


Cluster age was derived using PARSEC non-rotating isochrones \citep{tang2014new,chen2015parsec} at Z$\sim$0.006. These isochrones were used as they have the added advantage of coming with photometry. We used $M_{\rm initial}$ as a constraint and found an age of 14Myrs. \cite{patrick2016chemistry} estimated the age of NGC 2100 to be 20 $\pm$ 5 Myr using SYCLIST stellar isochrones \citep{georgy2013grids} at SMC metallicity and at solar metallicity. The explanation for this difference in cluster age is due to the use of non-rotating isochrones in our study as it is known stellar rotation causes stars to live longer, and hence infer an older cluster age. When using rotating isochrones we found the same cluster age.

\subsection{Extinction}
We determined the extinction due to the dust wind from the ratio of the input and output flux at 0.55  $\micron$. This extinction is intrinsic to the circumstellar dust shell and is independent of any foreground extinction. Due to scattering within the dust wind the effect of extinction is small, see Table 2. As discussed by \cite{kochanek2012absorption}, enough light is scattered by the dust shell back into the line of sight of the observer so little flux is lost. In apparent contradiction, \cite{davies2013temperatures} derived extinctions for a sample of RSGs in the SMC and LMC of a few tenths of a magnitude. As the mass of the progenitor RSGs to the IIP SN are found from mass-luminosity relations, an extinction this high could have an effect on the mass calculation, causing them to be underestimated. 

We next fit isochrones to the CMD of our sample and by dereddening this it was possible to further estimate extinction present for the RSGs. We used a 14Myr PARSEC stellar evolutionary track isochrone \citep{tang2014new,chen2015parsec}. After adjusting the isochrone to a distance of 50kpc and the extinction law towards the LMC \citep{koornneef1982gas} we found that there is additional extinction towards the RSGs that is not present for blue supergiants (BSGs) in the cluster, see Fig. \ref{fig:colcol}. The isochrone shows that the RSGs require additional extinction in order to fit with the model, with stars \#1 and \#2 possessing even further extinction (see Section 4.1). This is further to the foreground extinction already known to be present for NGC 2100 \citep[around 0.5 mag, ][]{niederhofer2015no}. From this we can infer an intrinsic RSG extinction of approximately $A_V$$\sim$0.5 that is not present for other stars in the cluster. 

We considered the possibility that this extra extinction could be due to cool dust at large radii from the stars not detectable in the mid-IR. To do this we created DUSTY models at 30K with an optical depth of 2, large enough to produce the extra extinction of $A_V$  of $\sim$0.5 mags. If this dust were present it would emit at around 100$\micron$ with a flux density of > 1Jy.  A flux this high would be within the detection limits for surveys such as Herschel's HERITAGE survey \citep{meixner2013herschel}, which mapped the SMC and LMC at wavelengths of 100$\micron$ and above. After checking this data we found no evidence of the stars within NGC2100 emitting at this wavelength, suggesting that the additional extinction local to RSGs is not caused by a spherically symmetric cold dust shell. 

We also considered the effect of differential extinction on the cluster. \cite{niederhofer2015no} found that a low level of differential extinction is present in NGC 2100, but after analysing Herschel 100$\micron$ to 250$\micron$ images \citep{meixner2013herschel} it seems the core of the cluster, where the RSGs are, remains clear of dust. Star \#2 is spatially coincident with the BSGs, whereas star \#1 is away from the cluster core. From Herschel images we see no reason to expect that the foreground extinction should be unusually high for these objects. We therefore see no argument for the RSGs having different foregound extinction than the BSGs. Clumpy cold dust at larger radii could potentially explain the extra extinction in RSGs, we investigate this possibility further in Section 4.1.

\subsection{Sensitivity to grain size distribution}
To check how robust our results were to a change in the grain size distribution, we created grids of models for various constant grain sizes of 0.1  $\micron$, 0.2  $\micron$, 0.4  $\micron$ and 0.5  $\micron$ (in addition to the 0.3  $\micron$ grain size). The maximum grain size of 0.5  $\micron$ was chosen as 0.5  $\micron$ was recently found to be the average grain size for dust grains around the well known RSG VY Canis Majoris \citep{scicluna2015large}. We then derived \mdot\ values. 

\begin{figure}
  \caption{Plot showing \mdot\ derived at each constant grain size. Each colour represents a different star from NGC2100. The stars chosen are representative of high \mdot\ (\#2), intermediate \mdot\ (\#7) and low \mdot\ (\#9).}
  \centering
  \label{fig:gsvsmdot}
    \includegraphics[width=\columnwidth]{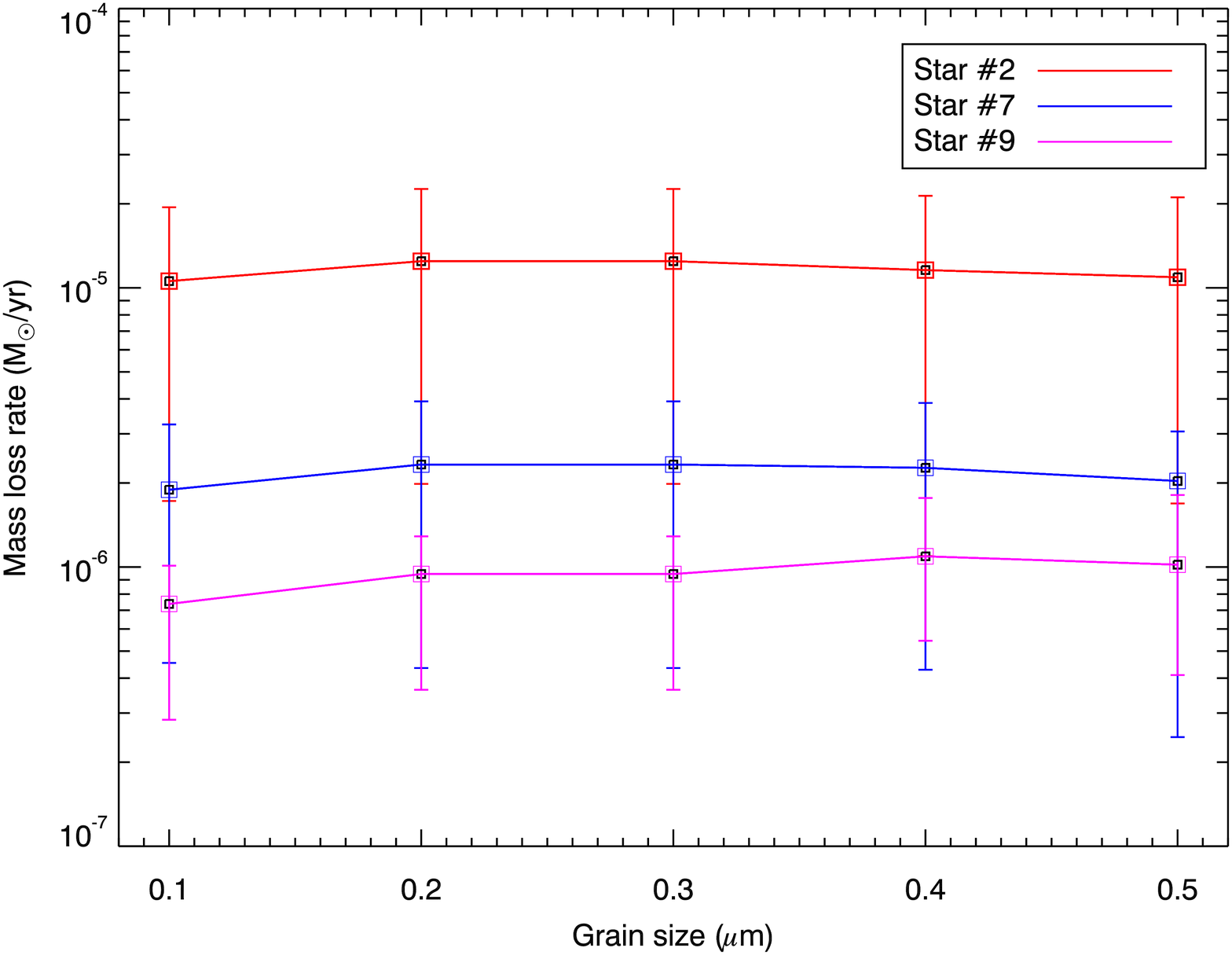}
\end{figure}

\begin{figure}
  \caption{Plot showing grain size versus A$_{v}$. Each colour represents a different star from NGC 2100. The stars chosen are representative of high \mdot\ (\#2), intermediate \mdot\ (\#7) and low \mdot\ (\#9).}
  \centering
  \label{fig:gsvsav}
    \includegraphics[width=\columnwidth]{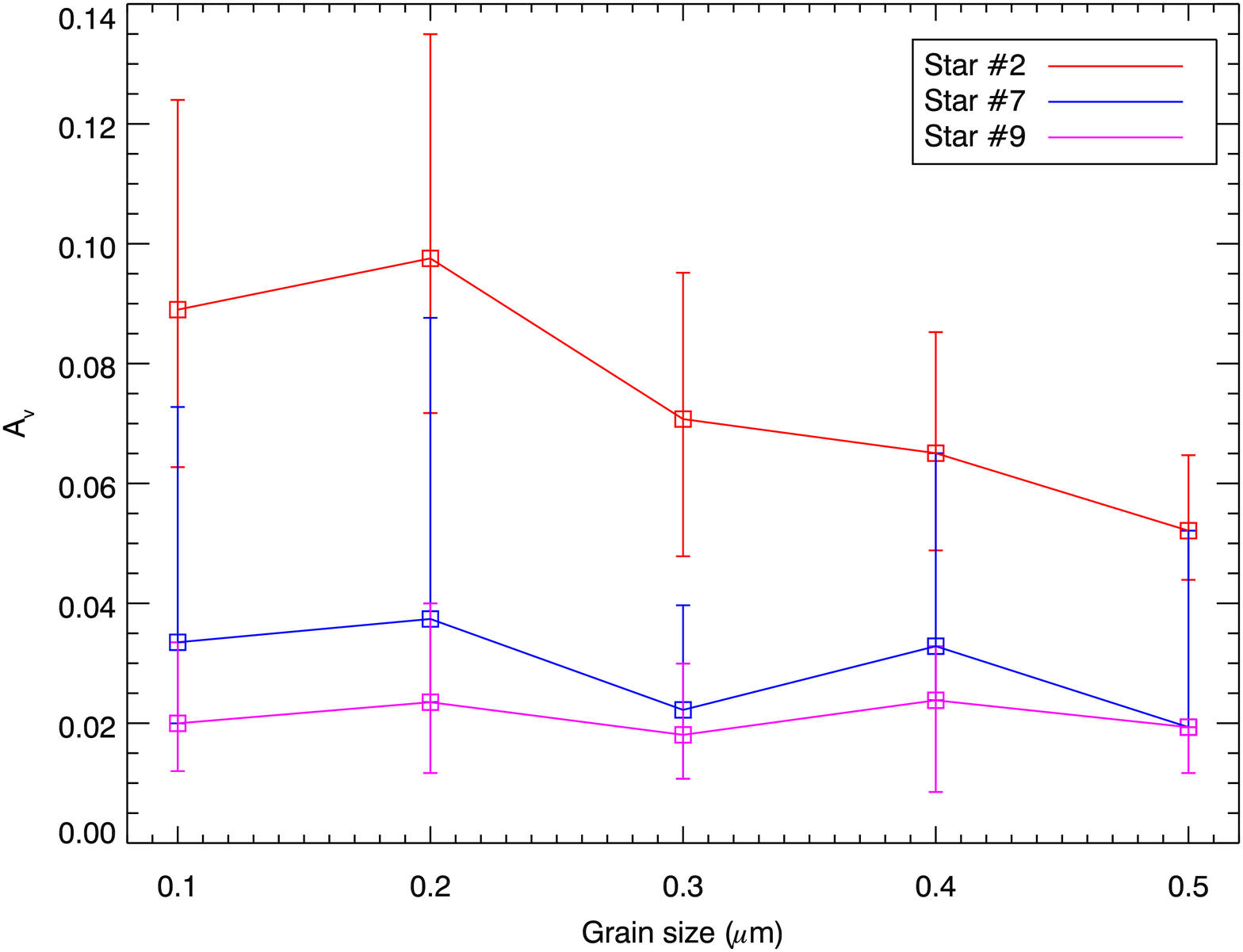}
\end{figure}

The results can be seen in Fig. \ref{fig:gsvsmdot}, where the \mdot\ value for each star is plotted for each constant grain size. It is clear that increasing grain size does not have an effect on the derived \mdot\ values. Similarly, the grain size also does not seem to have a significant effect on the value of $A_V$. This can be seen in Fig. \ref{fig:gsvsav}. The stars chosen are representative of high \mdot\ (\#2), intermediate \mdot\ (\#7) and low \mdot\ (\#9).While the $A_V$ does fluctuate slightly, the values remain within error boundaries of each other. A$_{v}$ is affected by grain size s described by Mie theory, which states that the scattering efficiency of the dust is dependent on grain size, $a$.  Extinction is dominated by particles of size $\sim$$\lambda/3$. When $\lambda$ $\gg$ grain size, scattering and absorption efficiency tend to 0. When dust grains are larger than a certain size, fewer particles are needed to reproduce the mid-IR bump causing a reduction in $A_V$. \mdot\ remains unaffected as the overall mass of the dust shell remains the same whether there are more smaller grains or fewer large grains. 

\section{Discussion}
\subsection{Evidence for enhanced extinction to stars \#1 and \#2}
As discussed in Section 3.3, stars \#1 and \#2 were found to be the most luminous in our sample, as well as having the strongest measured \mdot\ values. From our fitting procedure we found that the near-IR flux for these stars was overestimated by our best fit model spectrum. We now discuss possible causes for the discrepant near-IR photometry of these stars.

Next, we included the near-IR photometry into our fitting procedure to see the effect this would have on the output $L_{\rm bol}$. The $L_{\rm bol}$ values were derived by integrating under the best fit model spectrum for each star. As we did not initially include JHK photometry in our fitting procedure, it was possible that the derived $L_{\rm bol}$ would be overestimated also. When including the JHK photometry, the fits were improved at near-IR wavelengths but the mid-IR photometry fits became poorer. This had little effect on the best fit \mdot\ values and the trend of increasing \mdot\ with luminosity was still observed, well modelled by the dJ88 prescription \citep{de1988mass}, with stars \#1 and \#2 having lower $L_{\rm bol}$ by $\sim$0.3 dex. As we were unable to reliably fit both the near-IR and mid-IR photometry we concluded that stars \#1 and \#2 were intrinsically redder than the other, less evolved stars in our sample.

Choice of input SED could also have affected the measured $L_{\rm bol}$ as lower \teff\ causes the peak of the spectrum to shift to shorter wavelengths. The \teff\ of star \#1 has been found to be 4048$\pm$68K \citep{patrick2016chemistry} so we do not believe the underestimated near-IR flux to be an effect of input SED. Nevertheless, we repeated our fitting procedure using a lower \teff\ s of 3600K finding that this now underestimated the JHK photometry. We also calculated luminosities for each star based on the $K$-band calibration described by \cite{davies2013temperatures} and find the integrated luminosities of all stars are consistetent to a 1:1 relation within errors except for stars \#1 and \#2, which were underpredicted by this calibration. This further supports the suggestion that these stars have more self-extinction than the other RSGs in our sample. 

After repeating our fitting procedure with lower \teff\ and to include near-IR photometry we believe the most likely explanation is that stars \#1 and\#2 have extra extinction that cannot be explained by the inner dust wind. These stars are the most evolved in our sample, so it is possible this enhanced extinction only becomes apparent towards the end of an RSGs life. 

|t is known that RSGs have extended clumpy nebulae, for example $\mu$ Cep \citep{de2008red}. If $\mu$ Cep were at the distance of the LMC, the cold dust emitting at 100$\micron$ would be too faint to be observable, at a level of around 0.2 Jy (before we account for a factor of 2 lower dust to gas ratio for the LMC).  It is therefore possible that the enhanced extinction we observe for stars \#1 and \#2 is caused by the stars being surrounded by cold, clumpy dust that emits at similarly low levels.

We considered the possibility that the poor fits to the JHK and mid-IR photometry for stars \#1 and \#2 is due to extreme variability. If the mid-IR data we used in our analysis was taken at a time when the near-IR brightness of these stars was lower than when the 2MASS data were taken, this would cause our best fit SED to overestimate the flux at the JHK wavelengths. Star \#1 ($\equiv$HV 6002) is variable in the J and H bands by 0.13 mag and 0.11 mag, respectively (from minimum observed brightness to maximum observed brightness) and the 2MASS photometry we use in our analysis is the peak of this variability. In the V-band this variability is higher (~ 0.6 mag). We find that even at maximum brightness, the V-band photometry (corrected for foreground reddening) does not fit with our best fit SED. When we further de-redden the V-band photometry for the intrinsic reddening implied by the difference between our fit and the JHK photometry, we find the V-band photometry fits well with the best fitting SED with no tuning. We therefore conclude variability cannot explain the missing flux at JHK from our mid-IR photometry fits for stars \#1 and \#2. However, if we attribute this extra reddening to extinction, this could provide a self consistent explanation. 

\subsection{Effects of using a shallower density distribution}
\cite{shenoy2016searching} presented a recent study of cool dust around hypergiant stars VY Canis Majoris and $\mu$ Cep. Using photometry and DUSTY modeling to derive \mdot\ values, they adopted a fixed inner radius temperature of $T_{\rm in}$=1000K and a power law dust mass density distribution ($\rho$(r) $\propto$ r$^{-q}$) with a single index q throughout the shell. They then went on to test a range of optical depths and a range of power law indices q$\leq$2. They found that a power law with a q=2 did not produce enough cool dust to match the long wavelength end of the observed SED, instead concluding that a power law of $\rho$(r) $\propto$ r$^{-1.8}$ was more appropriate. This implies \mdot\ decreases with time since there was more dust present at large radii than would be for a fixed \mdot.

 By setting a fixed $T_{\rm in}$ at the sublimation temperature for silicate dust \cite{shenoy2016searching} are left with not enough cool dust at large radii. However, it is possible that the data could be fit equally well by fixing q=2 and allowing $T_{\rm in}$ to vary. We tested this for $\mu$ Cep by creating a model using the best fit parameter's found by \cite{shenoy2016searching} using the same density distribution ($T_{\rm in}$=1000K, $\tau_{37.2\micron}$=0.0029 and q=1.8) and then attempted to fit this spectrum using a q=2 density law and allowing $T_{\rm in}$ to vary. We found that a model with an inner dust temperature of 600K and q=2 density law fit Shenoy et al.'s model to better than $\pm$10\% at all wavelengths $\leq$70$\micron$, comparable to the typical photometric errors. If we include the 150$\micron$ data-point, noting that Shenoy et al.'s best-fit model overpredicted the flux of $\mu$ Cep at this wavelength. We can again fit the q=1.8 model with a steady state wind by adjusting the $T_{\rm in}$ value to 500K, giving a fit to better than 15\% at all wavelengths.

\cite{shenoy2016searching} also fit intensity profiles to the PSF of $\mu$ Cep. Models were computed using different density power law indices (q=1.8 and q=2) and a constant inner dust temperature of 1000K. Shenoy et al. concluded the PSF of $\mu$ Cep was best matched by an intensity profile of q=1.8 and $T_{\rm in}$ = 1000K out to 25 arcseconds. To check the robustness of this conclusion, we created DUSTY models using the model atmosphere in our grid most similar to that of Shenoy et al. (MARCS, $T_{\rm eff}$ = 3600K), with the same parameters as in Shenoy et al. (q=1.8, $T_{\rm in}$ = 1000K). We then also created a second DUSTY model using the parameters we found to give an equally good fit to the SED (q=2, $T_{\rm in}$ = 600K, discussed previously). The intensity profiles for both of these models was convolved with the PSF from Shenoy et al. We found the two models to be indistinguishable for both the SED and the intensity profile out to 25 arcseconds. From this we conclude that $\mu$ Cep data can be equally well modelled by a steady state wind and a cooler inner dust temperature. 

A density power law index q<2 implies a mass-loss rate that decreases over time.  Specifically, if $R_{\rm out}$ = 1000$R_{\rm in}$ then \mdot\ will be found to decrease by a factor of 1000$^{2 - q}$ in the time it takes for the dust to travel from $R_{\rm in}$ to $R_{\rm out}$. For q=1.8, \mdot\ would decrease by a factor of 4 through the time it takes for the dust to travel to the outer radius. In the case of $\mu$ Cep, \cite{shenoy2016searching} concluded that the \mdot\ must have decreased by a factor of 5 (from $5 \times 10^{-6}$ to $1 \times 10^{-6}$ M$_\odot$ yr$^{-1}$) over a 13,000 year history. If $\mu$ Cep's \mdot\ increases as the star evolves to higher luminosities, as we have found for the RSGs in NGC 2100\footnote{Although $\mu$ Cep has a higher initial mass and metallicity compared to NGC 2100, all evolutionary models predict an increase in luminosity with evolution, with the length of the RSG phase depending on the mass loss.} then this is inconsistent with the conclusions of Shenoy et al. This inconsistency can be reconciled if we assume the winds are steady-state (q=2) and allow $T_{\rm in}$ to be slightly cooler. From our best fit q=2 model we find an \mdot\ value of $3.5 \times 10^{-6}$, corresponding approximately to a density-weighted average of Shenoy et al.'s upper and lower mass loss rates.
 
As a further test of our conclusions that \mdot\ increases with evolution, we ran our fitting procedure and this time set our $T_{\rm in}$ to a constant value of 1200K. We still find an increase in \mdot\ with evolution. Although the fits at this constant $T_{\rm in}$ are worse at longer wavelengths, the warm dust (i.e. the most recent ejecta) is still accurately matched at shorter wavelengths (<8$\micron$). This relative insensitivity of \mdot\ to the inner dust radius is illustrated in Fig. \ref{fig:allcont}, where the contours of constant \mdot\ run parallel to the $\chi^2$ trenches. This shows again the degeneracy of optical depth and $T_{\rm in}$ where many combinations of the two result in the same value of \mdot. Even when fixing $T_{\rm in}$, we still find a positive correlation between \mdot\ and luminosity. 

\subsection{Consequences for stellar evolution}
We find a clear increase in \mdot\ with RSG evolution, by a factor of $\sim$40 through the lifetime of the star. These results are well described by mass-loss rate prescriptions currently used by some stellar evolution models, particularly dJ88 which matches the \mdot\ of the most evolved RSGs in our study (see Fig. \ref{fig:mdotvsL}). We find very little spread of $L_{\rm bol}$ with \mdot\, unlike that observed for field RSGs \citep[e.g.][]{van2005empirical}. The spread observed in previous results could be due to a varying $M_{\rm initial}$ in the sample stars. By focussing our study on a coeval star cluster we have kept metallicity and initial mass fixed, showing the mass-loss rate prescriptions fit well for LMC metallicity and $M_{\rm initial}$ of 14M$_\odot$. 

Mass loss due to stellar winds is a hugely important factor in determining the evolution of the most massive stars. There is uncertainty about the total amount of mass lost during the RSG phase, and therefore about the exact nature of the immediate SNe progenitors. \cite{meynet2015impact} studied the impact of \mdot\ on RSG lifetimes, evolution and pre-SNe properties by computing stellar models for initial masses between 9 and 25M$_\odot$ and increasing the \mdot\ by 10 times and 25 times. The models were computed at solar metallicity (Z$\sim$0.014) for both rotating and non-rotating stars. It was found that stronger \mdot\ had a significant effect on the population of blue, yellow and RSGs. It has been discussed previously that yellow supergiants (YSGs) could be post-RSG objects \cite[e.g.][]{georgy2012yellow,yoon2010evolution}, suggesting a possible solution to the "missing" Type IIP SNe progenitors. \cite{georgy2015mass} also discuss the case for an increased \mdot\ during the RSG phase. By increasing the standard \mdot\ by a factor of 3 in the models, \cite{georgy2015mass} find a blueward motion in the HRD is observed for stars more massive than 25M$_\odot$ (non-rotating models) or 20M$_\odot$ (rotating models, see \cite{georgy2012yellow}. 

As can be seen in Fig. \ref{fig:mdotvsL} we find the accepted \mdot\ prescriptions commonly used in stellar evolution codes fit well when the variables Z and $M_{\rm initial}$ are fixed. For this $M_{\rm initial}$ ($\sim$15M$_{\odot}$) and at LMC metallicity altering the \mdot\ prescriptions seems unjustified. Increasing the \mdot\ by a factor of 10 \citep[as in][]{meynet2015impact} would result in a strong conflict with our findings. 

We plan to further study this by looking at higher mass RSGs in galactic clusters at solar metallicity. This will allow us to  make a better comparison to the evolutionary predictions discussed by \cite{meynet2015impact} and \cite{georgy2015mass}. As well as this, the type IIP SNe that have been observed have all been of solar metallicity so it will be possible to make more accurate comparisons. 

\subsubsection{Application to SNe progenitors and the red supergiant problem}
In the previous sections, we have found that the most evolved stars in the cluster appear more reddened than others within the cluster. We now ask the question, if star \#1 were to go SN tomorrow, what would we infer about it's initial mass from limited photometric information? This is relevant in the context of the "red supergiant problem", first identified by \cite{smartt2009death} and updated in \cite{smartt2015observational}. Here it is suggested that RSG progenitors to Type IIP SNe are less massive than predicted by stellar evolution theory. Theory and observational studies strongly suggest that the progenitors to Type II-P events are red supergiants (RSGs) and could be anywhere in the range of 8.5 to 25M$_{\odot}$ \citep[e.g.][]{meynet2003stellar,levesque2005physical}. However, no progenitors appeared in the higher end of this predicted mass range, with an upper limit of 18M$_{\odot}$. Many of the luminosities (and hence masses) in this study were based on upper limit single band magnitudes only. In each \cite{smartt2009death} assumed a spectral type of M0 ($\pm$ 3 subtypes) and hence a BC$_v$ of -1.3 $\pm$ 0.3. The level of extinction considered was estimated from nearby stars or from Milky Way dust maps \citep{schlegel1998maps}. The presence of enhanced reddening that may occur at the end of the RSGs life, such as we observe for the two most evolved stars in our study (\#1 and \#2), was not considered.

We now apply similar assumptions to those of \cite{smartt2009death} to \#1, to see what we would infer about the star's initial mass were it to explode tomorrow. We find an excess reddening between J-K of 0.2 and an excess between H-K of 0.15, assuming \teff\ = 3900K. If we attribute this reddening to extinction, this implies an average $K_s$-band extinction of $A_K$ = 0.23 $\pm$ 0.11, leading to an optical $V$-band extinction of $A_V$ = 2.1 $\pm$ 1.1 \citep[based on LMC extinction law][]{koornneef1982gas}. If we take this stars' measured V-band magnitude \citep[$m_V$ = 13.79][]{bonanos2009spitzer} and adjust to $m_{\rm bol}$ using the bolometric correction BCv = -1.3 \citep[in line with][]{smartt2009death}, the measured L$_{\rm bol}$ without considering any extra extinction is 10$^{4.33}$L$_\odot$. When we factor in the extra reddening, this increases to 10$^{5.14 \pm 0.44}$L$_\odot$, in good agreement with the luminosity we derived from integration under the best fit DUSTY spectra. This increase will have a significant effect on the mass inferred. When extinction is {\it not} considered a mass of 8M$_\odot$ is found. From mass tracks, we have determined the initial mass of the cluster stars to be in the range of 14M$_\odot$-17M$_\odot$. Hence, the mass determined for the most evolved star in the cluster from single band photometry is clearly underestimated when applying the same assumptions as used by \citep{smartt2009death}. When extinction {\it is} taken into account the mass increases to $\sim$17 $\pm$ 5 M$_\odot$ (in close agreement with the mass inferred from cluster age, see Section 3.2). 

An alternative explanation for the redder colours of \#1 and \#2 is that they may have very late spectral type. Indeed, spectral type has been speculated to increase as RSGs evolve \citep{negueruela2013population,davies2013temperatures}. A colour of (J-K) = 0.17 would imply a supergiant of type M5 \citep{koornneef1983near,elias1985m}. If we consider stars \#1 and \#2 to be of this spectral type, this would require a BC$_v$ of approximately -2.3, giving a luminosity of $\sim$ 10$^{4.73} L_{\odot}$. This would lead to an inferred mass of 11 M$_\odot$, an increase on the 8$M_\odot$ inferred when the star was assumed to be of type M0,  but still lower than the 14M$_\odot$ - 17M$_\odot$ found from mass tracks.  

Based on the enhanced reddening we have observed for stars \#1 and \#2 it is interesting to see what effect an increased level of extinction would have on other progenitors studied by \cite{smartt2009death}. We considered three case studies, the progenitors to SN 1999gi, 2001du and 2012ec (of which SN 1999gi and 2001du are based on upper limits, with SN 2012ec having a detection in one band). We have chosen these SNe as they have host galaxies with sub-solar metallicity comparable to the LMC. 

\begin{itemize}
\item{SN 1999gi}

The progenitor site to SN 1999gi was first studied by \cite{smartt2001upper}, the 3$\sigma$ detection limit was determined to be $m_{F606W}$ = 24.9 leading to a luminosity estimate of log($L_{\rm bol}$/$L_\odot$)$\sim$4.49 $\pm$ 0.15 and upper mass limit of 14M$_\odot$. The upper limit to this luminosity was revisited by \citep{smartt2015observational} and revised upwards to be 10$^{4.9}$  once an ad-hoc extinction of $A_V$ = 0.5 was applied. Based on STARS and Geneva models, \cite{smartt2015observational} find the upper limit to the progenitor star's initial mass to be 13M$_\odot$. If we assume the progenitor to SN 1999gi had similar levels of extinction to star \#1 ($A_V$ = 2.4, including the ad-hoc extinction applied by Smartt). This leads to an extra $R$-band extinction $A_R$ = 1.4 \citep{koornneef1982gas} and therefore an increase in luminosity of 0.58 dex. This revises the upper limit on initial mass to 23 M$_\odot$, substantially higher than the upper mass originally stated. 

\item{SN 2001du}

This RSG progenitor was observed in the F336W, F555W and F814W bands, which were all non detections. The 3$\sigma$ upper limit was based on F814W as this waveband is least affected by extinction.  From this \cite{smartt2009death} find a luminosity of log($L_{\rm bol}$/$L_\odot$)$\sim$4.57 $\pm$ 0.14. When including an extra ad hoc $A_V$ = 0.5, \cite{smartt2015observational} find the mass of this progenitor to be 10M$_\odot$ and a luminosity of 10$^{4.7}$$L_{\rm bol}$. If we again assume additional optical extinction $A_V$ = 1.4 \cite[on top of the ad hoc extinction included by ][]{smartt2015observational} we find an $I$-band extinction $A_I$ = 0.95 leading to an increase in measured $L_{\rm bol}$ of 0.38 mag. This would revise the upper mass limit for this progenitor to $\sim$ 17 M$_\odot$. 

\item{SN 2012ec}

Finally the RSG progenitor to SN 2012ec, originally discussed by \cite{maund2013supernova}. These authors used a foreground reddening of E(B-V)=0.11 and constrained $T_{\rm eff}$ to < 4000K using an upper limit in the F606W band. Using a $F814W$ pre-explosion image the progenitor candidate is found to have a brightness of $m_{F814W}$ = 23.39 $\pm$ 0.18. \cite{maund2013supernova} estimate the luminosity to be log($L_{\rm bol}$/$L_\odot$) = 5.15 $\pm$ 0.19 leading to a mass range of 14 - 22 M$_\odot$. If we again apply a similar level of extinction we measure for star \#1 to the progenitor of SN 2012ec we infer a luminosity of log($L_{\rm bol}$/$L_\odot$) = 5.41, leading to a mass of between 22 - 26 M$_\odot$ based on Fig. 2 of \cite{smartt2009death}.

\end{itemize}
From the three case studies above, we have shown that by including similar levels of reddening that we find in the most evolved stars in NGC 2100, the initial mass estimates for Type IIP SN progenitors increase substantially. When applied to all objects in the \cite{smartt2009death} sample this may resolve the inconsistency between theory and observations and hence solve the red supergiant problem.

One argument against extinction being the cause of the red supergiant problem comes from X-ray observations of SN. \cite{dwarkadas2014lack} used the X-ray emission from IIP SNe to estimate the pre-SNe \mdot\ for RSGs, arguing for an upper limit of 10$^{-5}$M$_\odot$yr$^{-1}$. By using the mass loss rate - luminosity relation of \cite{mauron2011mass} and the mass-luminosity relation from STARS models \citep{eggleton1971evolution}, this upper limit to the mass-loss rate was transformed into an upper mass limit of 19M$_\odot$, in good agreement with \cite{smartt2009death}. While this number is in good agreement with \cite{smartt2009death} we estimate the errors on this measurement must be substanstial. \cite{dwarkadas2014lack} converts an X-ray luminosity into a value of \mdot\ (a conversion which must have some systematic uncertainties, but as we do not know we assume this to be consistent with zero) and from this \mdot\ finds a luminosity of the progenitor using the calibration in \cite{mauron2011mass}. This calibration between \mdot\ and luminosity has large dispersion of a factor of ten (see Fig. 5 of \cite{mauron2011mass}), but if we are again optimistic take these to be half that, a factor of five. From this, a progenitor mass was calculated under the assumption from mass-luminosity relation for which RSG luminosity scales as L $\sim$ M$^{2.3}$, increasing the errors further. Even with our optimistic estimates, we find the error to be $\pm$\ a factor of two, or around 19$\pm$10M$_\odot$. Therefore, we conclude that X-ray observations of IIP SNe provide only a weak constraint on the maximum initial mass of the progenitor star, and cannot rule out that circumstellar extinction is causing progenitor star masses to be underestimated.

\section{Conclusion}
Understanding the nature of the mass loss mechanism present in RSGs remains an important field of study in stellar astrophysics. Here a method of deriving various stellar parameters, $T_{\rm in}$, $\tau_V$, \mdot\, was presented as well as evidence for an increasing value of \mdot\ with RSG evolution. By targetting stars in a coeval cluster it was possible to study \mdot\ while keeping metallicity, age and $M_{\rm initial}$ constrained. As all stars currently in the RSG phase will have the same initial mass to within a few tenths of a solar mass, it is possible to use luminosity as a proxy for evolution, due to those stars with slightly higher masses evolving through the HR diagram at slightly faster rates. From our study we can conclude the following:
\begin{itemize}
\item The most luminous stars were found to have the highest value of \mdot\, evidenced observationally by colour-magnitude diagrams and also by a positive correlation between bolometric luminosity and \mdot. 

\item Our results are well modelled by various mass-loss rate prescriptions currently used by some stellar evolution groups, such as dJ88 and Reimer's, with dJ88 providing a better fit for the RSGs with stronger \mdot. We therefore see no evidence for a significantly increased \mdot\ rate during the RSG phase as has been suggested by various stellar evolutionary groups

\item We also presented extinction values for each star, first determined from DUSTY models and next determined by isochrone fitting. While the warm dust created low extinction values in the optical range ($A_V$$\sim$0.01 mag), isochrone fitting showed that RSGs may have an intrinsic optical extinction of approximately $A_V$ = 0.5mag. This extinction cannot come from the warm inner dust, but may come from clumpy cool dust at larger radii. This supports the suggestion that RSGs create their own extinction, more so than other stars in the same cluster. 

\item We also find that the two most luminous (therefore most evolved) stars in our sample show enhanced levels of reddening compared to the other RSGs. If we attribute this reddening to further extinction, this implies an average $K_S$-band extinction of $A_K$ = 0.23 $\pm$ 0.11. We do not find evidence for cold dust emitting at wavelengths of 100$\micron$ as we first suspected, so we as yet do not know the source of this extra reddening towards the RSGs.

\item When taking the enhanced reddening into account it seems the inferred progenitor masses to Type II-P SNe often increase significantly, providing a potential solution to the red supergiant problem. If this level of extinction is applied to all known RSG progenitors (assuming all RSGs show enhanced reddening at the end of their lives) the inconsistency between theory and observations may be resolved. 
\end{itemize}
 Future work will involve applying this technique to RSGs at solar metallicity to see if the mass-loss rate prescriptions are still appropriate. We also plan to apply this technique to clusters where the stars have higher initial masses closer to the upper RSG limit. 
\section*{Acknowledgements}
We thank the anonymous referee for comments which have helped us improve the paper. We thank Rolf-Peter Kudritzki and Stephen Smartt for useful discussions. We also acknowledge the use of the SIMBAD database, Aladin, IDL software packages and astrolib. 




\bibliographystyle{mnras}
\bibliography{references} 





\bsp	
\label{lastpage}
\end{document}